\documentclass{article}
\usepackage{lmodern}
\usepackage{amssymb,amsmath}
\usepackage{ifxetex,ifluatex}
\usepackage{fixltx2e} 
\ifnum 0\ifxetex 1\fi\ifluatex 1\fi=0 
  \usepackage[T1]{fontenc}
  \usepackage{hyperref}

\usepackage{mathtools}
\mathtoolsset{showonlyrefs}

\usepackage{graphicx}
\usepackage{subcaption}
\usepackage{xcolor}
\usepackage{bm}
\usepackage{amsmath}
\usepackage{bbm}
\usepackage{enumerate}

\usepackage[raggedright]{titlesec}
\titleformat{\paragraph}[hang]{\normalfont\normalsize\bfseries}{\theparagraph}{1em}{}
\titlespacing*{\paragraph}{0pt}{3.25ex plus 1ex minus .2ex}{0.5em}

\newcommand{\Ssb}{\bm{\mathsf{S}}}
\newcommand{\Esb}{\bm{\mathsf{E}}}

\newcommand{\usb}{\bm{\mathsf{u}}}
\newcommand {\zetab} {\bm{\zeta}}

\newcommand {\cb} {\mathbf{c}}

\newcommand {\eb} {\mathbf{e}}

\newcommand {\nb} {\mathbf{n}}

\newcommand {\tb} {\mathbf{t}}
\newcommand {\xb} {\mathbf{x}}
\newcommand {\yb} {\mathbf{y}}
\newcommand {\ub} {\mathbf{u}}

\newcommand {\wb} {\mathbf{w}}

\newcommand {\Eb} {\mathbf{E}}

\newcommand {\Kb} {\mathbf{K}}

\newcommand {\Ib} {\mathbf{I}}

\newcommand {\Sb} {\mathbf{S}}
\newcommand {\Tb} {\mathbf{T}}

\newcommand {\Co} {\mathbb{C}}
\newcommand{\dd}  {\mathrm{d}}  
\newcommand {\Ec}  {\mathcal{E}}
\newcommand {\Hc}  {\mathcal{H}}
\newcommand {\Uc}  {\mathcal{U}}
\newcommand {\Pc}  {\mathcal{P}}
\newcommand {\Sigmab}    {\mathbf{\Sigma}}
\def\div{\mathop{\hbox{div}}}

\date{}

\begin{document}
	\title{\textbf{Towards a phase-field fracture model for thin structures: a coarse-grained constitutive law for  beams}}
\author{Giovanni Corsi\footnote{	\href{mailto:giovanni.corsi@uniroma1.it}{giovanni.corsi@uniroma1.it}}, Antonino Favata\footnote{	\href{mailto:antonino.favata@uniroma1.it}{antonino.favata@uniroma1.it}}, Stefano Vidoli\footnote{	\href{mailto:stefano.vidoli@uniroma1.it}{stefano.vidoli@uniroma1}}}

\maketitle

\vspace{-1cm}
\begin{center}
	{\small
		Department of Structural and Geotechnical Engineering\\
		Sapienza University of Rome \\
		Rome, Italy
	}
\end{center}

\pagestyle{myheadings}
\markboth{G.~Corsi, A.~Favata, S.~Vidoli }
{A constitutive law for brittle fracture of beams}

\vspace{-0.5cm}
\section*{Abstract}
Damage gradient models approximate fracture mechanics using a modulation of the material stiffness. To this aim a single scalar field, the damage, is used to degrade as a whole the elastic energy. If applied to the structural models of beams and shells, where the elastic energy is the sum of the stretching and bending contributions, a similar approach  is not able to capture some important features. For instance, the coupling between axial and bending strains induced by  cracks non-symmetric with respect to the center line is completely missed.
In this contribution, we deduce a   constitutive law for a beam having a crack non-symmetric with respect to the center line.  To achieve this, we  perform an asymptotic coarse-grained procedure from a 2D problem, using a sharp interface model. We deduce a homogenized 1D elastic energy, coupling the axial and bending strains, the constitutive parameters of which depend in a different way on the crack depth and we state how precisely they do. This should pave the way to a rational development of a phase-field gradient model for thin structures.

%

\section{Introduction}
This section is organized as follows: in Sec. \ref{phasefieldm} we first review  the relevant literature on phase-field models for thin structures; based on the numerical evidence shown  in Sec. \ref{2Dres} in a 2D framework, we anticipate  in Sec. \ref{results} our main results.

\subsection{Phase-field models for brittle fracture of thin structures}\label{phasefieldm}

In the literature there are different approaches to model fracture in shells, plates, and beams, such as mesh-free methods \cite{Amiri_2014,Weidong_2020,Yu_2020}, XFEM methods \cite{Areias_2005,Dolbow_2000,Nguyen_2015}, etc.

In solid mechanics, a breakthrough in modeling brittle fracture is the formulation of the Griffith theory in a variational framework \cite{Francfort_1998} and its  regularization through the Ambrosio-Tortorelli results \cite{Ambrosio_1990}, in  the so-called \textit{phase-field}  gradient models \cite{Bourdin_2000,Miehe_2010}.

These models are based on the minimization of an energy functional depending on both the displacement and  damage fields; the latter  --- say $\alpha$ ---  is introduced to approximate the sharp crack discontinuity in a domain with a smooth transition between the intact  and the fully broken material. A detailed discussion on the extensive literature on the subject is out of the scope of this paper; the interested reader is referred to some comprehensive reviews on the subject (see, \textit{e.g.}, \cite{Ambati_2015}). More specifically, within this framework, the energy functional to be minimized is given by
\begin{equation}\label{phasefield}
	\Ec(\ub,\alpha)=\frac12\int_{\Omega}\Co(\alpha)\Eb(\ub)\cdot\Eb(\ub)\,\dd\xb+ G_c\int_{\Omega}\left(\frac{(1-\alpha)^2}{4\delta}+\delta|\nabla\alpha|^2\right)\dd\xb,
\end{equation}
where $\ub$ is the displacement field, $\Eb(\ub)$ the corresponding strain, $\Co(\alpha)$ the elastic tensor, modulated by the phase-field $\alpha$,   $\delta>0$ a parameter that controls the width of the transition zone of $\alpha$, and $G_c$ the fracture toughness. Typically the elasticity tensor is assumed as a multiplicative function of the type $\Co(\alpha)=m(\alpha)\Co_o$. If the modulation function $m(\alpha)$ is chosen as $(1-\alpha)^2$, it has been proven that the functional \eqref{phasefield}, for $\delta\to 0$, approximates in the sense of $\Gamma$-convergence the energy functional of the Griffith theory of brittle fracture:
\begin{equation}\label{griffith}
	\Ec_G(\ub,\Gamma)=\frac12\int_\Omega \Co\Eb\cdot\Eb\,\dd\xb+G_c\int_{\Gamma}\dd s,
\end{equation} 
where  $\Gamma\subset\Omega$ is the crack set and the displacement $\ub$ is discontinuous across $\Gamma$.

This theory has been  adapted to model damage and crack in plates and shells \cite{Amiri_2014,Kiendl_2016,Kikis_2021}. In particular, in \cite{Kiendl_2016}, the functional \eqref{phasefield} is modified as follows:
\begin{equation}\label{DeLor}
\Ec_S=\int_A \Big(m(\alpha)\Psi_e^+(\Eb_m,\Kb)+\Psi_e^-(\Eb_m,\Kb)\Big)\dd A+ G_c\int_{A}\left(\frac{(1-\alpha)^2}{4\delta}+\delta|\nabla\alpha|^2\right)\dd A,
\end{equation}
where $A$ is the middle surface of the shell, $\Psi_e$ is the elastic energy, depending on the membrane deformation $\Eb_m$ and the curvature $\Kb$ of $A$; the superscript $\pm$ in $\Psi_e$ reflects the assumption that the total strain  is decomposed in a tensile and compressive contribution, in order to model  when the material cracks in tension but not in compression.

The strongest assumption in \eqref{DeLor}  is that the same modulation function $m(\alpha)$ affects both the membrane and bending energy contributions. As we will see, in very common circumstances such an assumption may lead to nonphysical results.  There are very few contributions in which this point is faced, briefly reviewed here. In \cite{Areias_2016}  two different phase-fields are considered, one for the top and the other for the bottom of the shell, so that it is possible to allow a differential crack between the thickness.  In \cite{Lai_2020} a double sigmoid Ansatz is introduced for the phase field along the thickness, allowing for a partial description of the evolution within the transversal section of a beam; in this contribution, as the damage Ansatz is symmetric with respect to the center line, the coupling between stretching and bending vanishes. In \cite{Mareddy_2022} a ``mixed-dimensional'' model is introduced, which combines structural elements representing the displacement field in the two-dimensional shell midsurface with continuum elements describing a crack phase-field in the three-dimensional solid space. In \cite{Brunetti_2020} the authors propose a subdivision of the plate-like three-dimensional solid into several layers, allowing to describe the evolution of the damage through the thickness; in this way, while the mechanical behaviour of the solid is governed by the classical theories of plates and shells, the phase fields equation has to be satisfied within each layer.

\begin{figure}[h!]
	\centering
	\includegraphics[scale=.5]{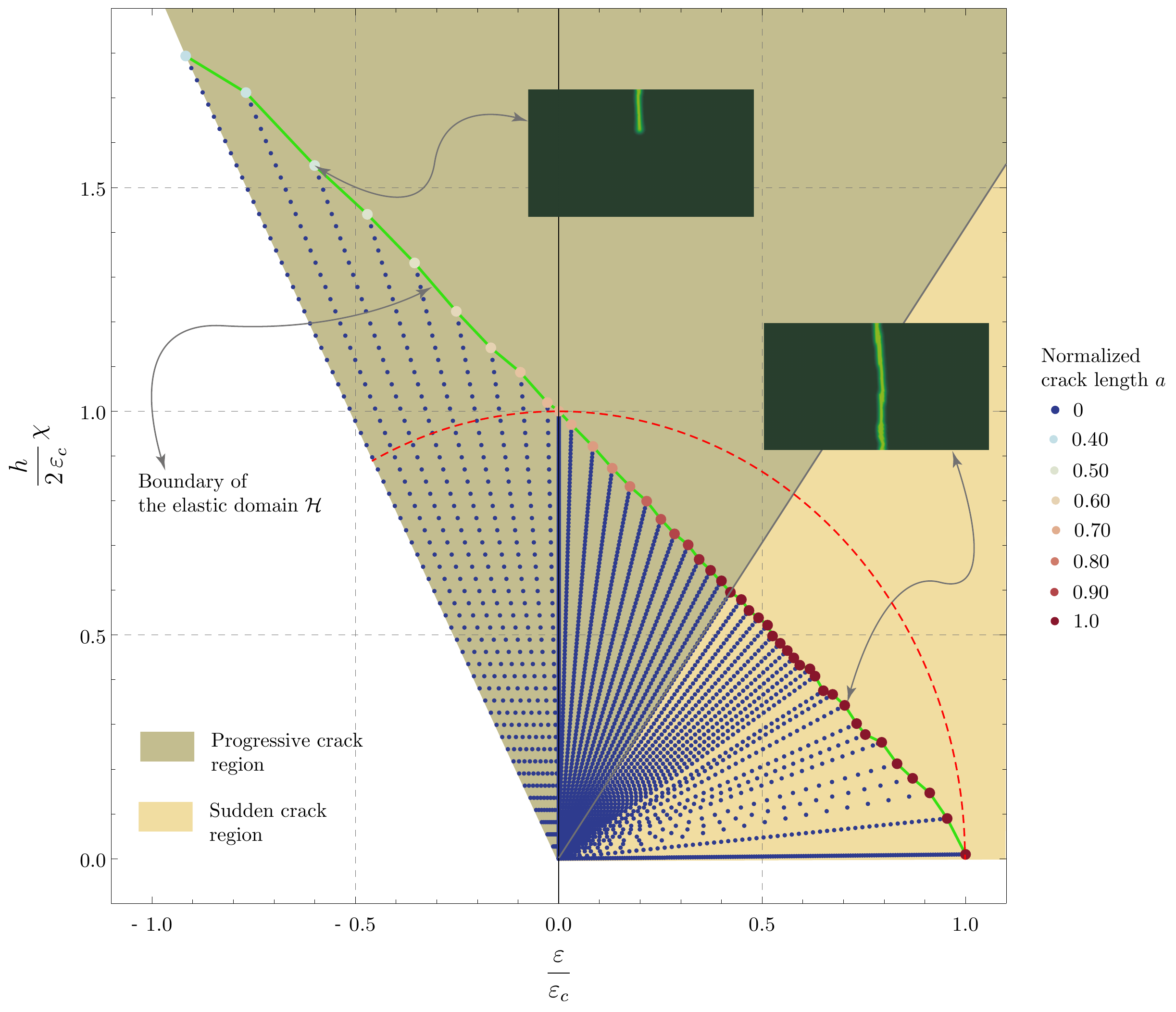}
	\caption{Damage scenario for a 2D rectangular domain subject to a combination of axial and bending strains. 
		}
	\label{fig:epschidiagram}
\end{figure}

\subsection{Numerical results in 2D}\label{2Dres}
In order to go towards a phase-field formulation of a 1D model for brittle fracture of beams, we find it instrumental to present some results borrowed from a  2D model. This will help to highlight the relevant features that one would expect to recover in a 1D beam model.

Let us consider a rectangular beam-like domain subject to a given displacement at the bases. More specifically,  let $\{x_1, x_2, x_3 \}$ denote the coordinate of a Cartesian reference system, and let a planar rectangular beam be  identified with the domain
\begin{equation}\label{domain}
\Omega=\{(x_1, x_, x_3)\,|\, -~L/2\leq x_1\leq L/2, -h/2\leq x_2\leq h/2,   -b/2\leq x_3\leq b/2 \},
\end{equation}
with  \(h < L < b\). 
The displacement $u_1$ in direction $x_1$ is fixed at the bases:
\begin{equation}
	u_1(\pm L/2,x_2)=\mp \frac{\Delta\vartheta}{2} x_2\pm \frac{\Delta l}{2},
\end{equation}
where $\Delta\vartheta$ and $\Delta l$ are incremental control parameters, meaning the relative rotation and axial displacement between the bases $x_1=\pm L/2$. The body is then subject to axial and bending strains.
We solve the problem numerically,  adopting the model \eqref{phasefield}, with $\Co(\alpha)=(1-	\alpha)^2\Co_o$, where $\Co_o$  is the elasticity tensor of an isotropic material, having Young modulus $Y$ and Poisson ratio $\nu$.

 We have used an incremental energy minimization framework \cite{Petryk_2003} where the imposed displacement is discretized in steps and  the total energy is minimized with respect to the displacement $\ub$ and phase field $\alpha$. 
Specifically,  an alternate minimization strategy has been adopted:  at each step a sequence of  minimizations with respect to $\ub$ at fixed $\alpha$ and minimizations with respect to $\alpha$ at a fixed $\ub$ is performed until convergence. 
While the former requires to solve a linear elasticity problem, the latter demands the  solution of a nonlinear problem (a modified Newton-Raphson scheme is used). Nevertheless, the minimization in $\alpha$ has the advantage to be local in space and, therefore, effectively parallelizable. The domain has been discretized with an unstructured mesh. The displacement and damage fields have been  taken in a piece-wise affine finite element space ($\mathcal{P}_{1}\Lambda^{0}(\Delta_2)$ \cite{Cockburn_2017}) over the domain. 
 The time has been discretized with non-uniform  steps. The code has been written in \texttt{python} as interface to \href{https://fenicsproject.org/}{\texttt{FEniCS}}, a popular open-source computing platform for solving partial differential equations \cite{Logg_2012,Aln_2015}. In particular the package \texttt{DOLFIN} \cite{Logg_2010}, a library aimed at automating solution of PDEs using the finite element method, has been used extensively. 

As output, we determined the damage field $\alpha(x_1, x_2)$, for given axial strain $\varepsilon=(\Delta l)/L$ and curvature $\chi=(\Delta \vartheta)/L$.
 As expected, the damage localizes into a small region, whose size is proportional to the characteristic length $\delta\ll L$; see the insets of Fig. \ref{fig:epschidiagram} for examples of such localizations.

In Fig. \ref{fig:epschidiagram} we also show the boundary $\Hc$ of the elastic domain  in the plane
$(\varepsilon,\chi)$. This boundary corresponds to the loss of stability of the solution $\alpha=0$ in $\Omega$, during a load process, monotonically increasing from $\varepsilon=\chi=0$, and having fixed the ratio between axial and bending strains. The colour of the points of the boundary represents  the normalized crack depth $a$, being  the actual crack depth $ah$; it is determined by the stable solution for $\alpha$ in an additional loading step. This allows to distinguish if the propagation of the crack is sudden or not. The axial strain and curvatures are scaled with respect to $\varepsilon_c$ and $(2\varepsilon_c)/h$, where 
 $\varepsilon_c$ is the \textit{critical strain}: 
 	\begin{equation}
 	\varepsilon_c=\sqrt{\frac 38\frac{G_c}{\delta}\frac{1-\nu^2}{Yb}},
 \end{equation}
 see \cite{Tanne_2018}.


Here the relevant features to be highlighted:
\begin{enumerate}
	\item The boundary $\Hc$ of the elastic domain  is a straight line, up to numerical errors. This is expected, as this line is close to the set of points $(\varepsilon, \chi)$ where the strain of the  beam upper fiber, namely $\varepsilon+\frac{h}{2}\chi$, reaches the critical value $\varepsilon_c$.
%
%
	\item As soon as the the solution $\alpha=0$ in  $\Omega$ becomes unstable, stable solutions for the damage field are characterized by localization on strips, at the center of which $\alpha=1$ (see insets in Fig. \ref{fig:epschidiagram}). There are two possible cases: in one this strip develops through the whole thickness,  and corresponds to a complete fracture of the beam cross section. In the other, due to the stabilizing effect of compressed fibers within the cross section, the beam equilibrium remains stable even if the crack depth is smaller than the thickness ($a<1$). 
\item The numerical evidences shown in Fig. \ref{fig:epschidiagram}  are in contrast with the assumption made in \eqref{DeLor} on the form of the damage modulation of the elastic energy. To  verify this assertion, we limit to the  sector  $\chi\geq 0$ and $\varepsilon\geq -\chi h/2$, and assume a quadratic dependence of $\Psi_e^+(\Eb_m,\Kb)$ on the strains $\Eb_m=\varepsilon\,\eb_1\otimes\eb_1$ and $\Kb=\chi\,\eb_1\otimes\eb_1$. We necessarily get 
$\Psi_e^+=1/2\left(\mathsf{C_\varepsilon}\varepsilon^2+\mathsf{C_\chi}\chi^2\right)$, for two positive constants $\mathsf{C_\varepsilon}$ and $\mathsf{C_\chi}$; any coupling term, bilinear in $\varepsilon$ and $\chi$, must vanish due to the symmetry of the domain and the material with respect to the center line. Hence, using \eqref{DeLor}, the elastic boundary is the set of points where
$m'(0)\Psi_e^+(\varepsilon, \chi)=G_c/(2\delta)$, \textit{i.e.}, a portion of a canonic ellipse shown in Fig. \ref{fig:epschidiagram} as a dashed red curve. 
The 2D numerical evidences obtained using \eqref{phasefield}
show that in a pure bending (traction) test a superimposition of a positive small axial (resp. bending) strain reduces the elastic limit. These effects are physically sound, but are completely missed when assuming \eqref{DeLor}.
\end{enumerate}

\subsection{Main results}\label{results}
In the light of these evidences provided by fully 2D simulations, we can claim that a functional like \eqref{DeLor},  where stretching and bending are modulated as a whole, cannot properly capture the relevant features of the fracture evolution within a thin structure.  
Aimed at a phase-field model overcoming these limitations, we here  
answer the following  questions arising when a beam undergoes a crack:
\begin{enumerate}
\item how do the bending and axial strains interact?
\item is it possible to deduce a suitable 1D constitutive model able to describe the global response, in terms of gross descriptors typical of 1D models of beams?
\item how do the resulting stiffnesses depend on the crack depth?
\end{enumerate}
We resort to a sharp interface model instrumental to understand the key ingredients a putative phase-field must have.

Our work is based on the following main steps:
\begin{enumerate}
	\item Performing an asymptotic coarse-graining procedure, \textit{i.e.}, solving a succession of 2D elasticity problems, through asymptotic expansions, in terms of a smallness parameter, related to the crack depth.
	\item Finding the gross transmission conditions at the crack. In particular, we deduce a 1D model in which the presence of the crack is retained by means of suitable jump conditions in correspondence of the crack;\footnote{This result is consistent with \cite{almi_2021}, where the dimensional reduction from 3D is performed for the functional \eqref{griffith}, by means of $\Gamma$-convergence. It is there proven that the 3D  displacement field converges to the classical Kirchhoff-Love Ansatz and the crack set converges to the set of the jumps of the displacement and the rotation.} this is equivalent to introduce some springs (extensional, torsional and mixed), whose stiffnesses are determined in terms of the crack depth.
	\item Regularizing the transmission conditions, `spreading' the jumps in a small region with measure $\ell$, whose order of magnitude is the same as the height of the beam. This leads to a continuous energy depending on the (normalized) crack depth $a$:
	\begin{equation}
		\Uc^{\ell}(a) =\frac{1}{2} 
		\int_{-\frac{\ell}{2}}^{+\frac{\ell}{2}} \left(\mathsf{C_\varepsilon}(a) \varepsilon^2+2\mathsf{C_{\varepsilon\chi}}(a) \varepsilon\chi +\mathsf{C_{\chi}} (a)\chi^2 \right) \,\text{d} x_1,
	\end{equation}
and to the constitutive law
\begin{equation}\label{constlaw}
	N=\mathsf{C_\varepsilon}(a)\varepsilon+\mathsf{C_{\varepsilon\chi}}(a)\chi, \qquad 
	M=\mathsf{C_\chi}(a)\chi+\mathsf{C_{\varepsilon\chi}}(a)\varepsilon,
\end{equation}
which is not postulated, but rigorously deduced. The crucial point is how the stiffnesses depend on $a$. Since each coefficient depends on $a$ in a different manner (see Fig. \ref{fig:rigidezze} and eq. \eqref{stiffnesses}), we conclude that it is not possible to postulate the degradation function as  in  \eqref{DeLor}.  The overall response of the beam turns out to be consistent with the numerical simulation performed on a fully 2D phase-field model.
\end{enumerate}

\section{Asymptotic coarse-graining formulation}
In order to determine a constitutive law for a cracked beam, we carry out a coarse-graining formulation, based on asymptotic expansions. In \cite{Baldelli2021} a similar path is followed, but limiting attention to symmetric cracks. 

\begin{figure}
	\centering
	\includegraphics[scale=.8]{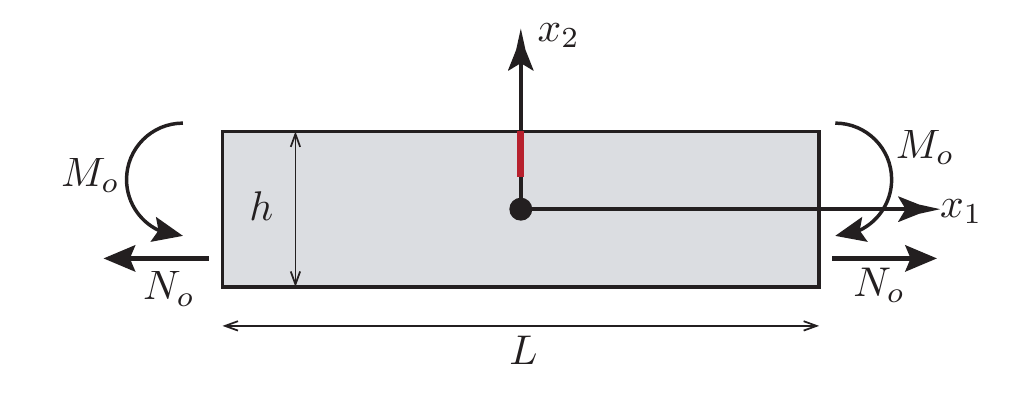}
	\caption{Geometry and applied loads.}
	\label{fig:geometria}
\end{figure}

  We consider a planar rectangular beam,  identified with the domain $\Omega$ in eq. \eqref{domain}, with  \(h \ll L \ll b\).  $\Omega$ is supposed to have a vertical straight crack stemming from the apical side of the beam (see Fig. \ref{fig:geometria}).

We introduce
the dimensionless parameter \(\eta = h/L\), and we will study the
elasticity problem, in a small displacements setting. Our
approach will be asymptotic with respect to the parameter \(\eta\), and
we will assume that the depth of the crack is of order \(\eta\).
Furthermore, we  assume the plane strain hypothesis, and
that the solid is loaded only at the bases \(x_1=\pm L/2\).
Let $\Sb=S_{ij}\eb_i\otimes\eb_j$ denote the Cauchy stress within the solid, and $N$, $M$ the gross stress measures of the beam, the\textit{ axial force} and the \textit{bending moment}:
\begin{equation}
	N(x_1)=b\int_{-h/2}^{h/2} S_{11}(x_1, x_2) \dd x_2 , \qquad M(x_1)=-b\int_{-h/2}^{h/2}
	x_2S_{11}(x_1, x_2) \dd x_2.
\end{equation}
We  assign the stress distribution at the boundary $x_1=\pm L/2$:
\begin{equation}
    S_{11}(\pm L/2, x_2) = \frac{N_o}{bh} - \frac{12 M_o x_2}{bh^3},
    \label{eq:bc_stress_distribution_dimensional}
\end{equation}
(see Fig. \ref{fig:geometria}).
The other boundaries of the domain are stress free. Furthermore a choice on the scaling of the
external forces with respect to the small parameter is necessary: we
assume that the moment and traction be related as follows
\begin{equation}
    M_o \propto N_o h,
    \label{eq:dimensional_scaling_assumption}
\end{equation}
that is, the moment is one order smaller than the traction with respect
to the small parameter. The equations of the problem are those of linear
elasticity:

\begin{equation}
\div \Sb = \mathbf{0}, \qquad 
    Y\Eb = - \nu \text{tr}(\Sb)\Ib+ (1+\nu) \Sb, \qquad
    \Eb = \text{sym}\nabla \ub,
    \label{eq:elasticity_dimensional}
\end{equation}
where \(\Eb\) is the linearized strain measure and  
\(\ub\) the displacement field. $Y$  and $\nu$ are the Young modulus and the Poisson coefficient, respectively. Within the plane strain assumption, we
have \(\Eb_{i3} = 0\), and from the second equation of
(\ref{eq:elasticity_dimensional}):

\begin{equation}
    S_{\alpha 3} = 0, \qquad S_{33} = \nu (S_{11} + S_{22}).
\end{equation}
Here and henceforth, Latin indices range from 1 to 3, and that the
Greek indices range from 1 to 2. In this context, the compatibility
equations reduce to the single equation:

\begin{equation}
    E_{11,22} + E_{22,11} = 2 E_{12,12}.
    \label{eq:compatibility_dimensional}
\end{equation}

\hypertarget{non-dimensionalization}{%
\subsection{Non-dimensionalization}\label{non-dimensionalization}}

\begin{figure}
	\centering
	\includegraphics[scale=0.7]{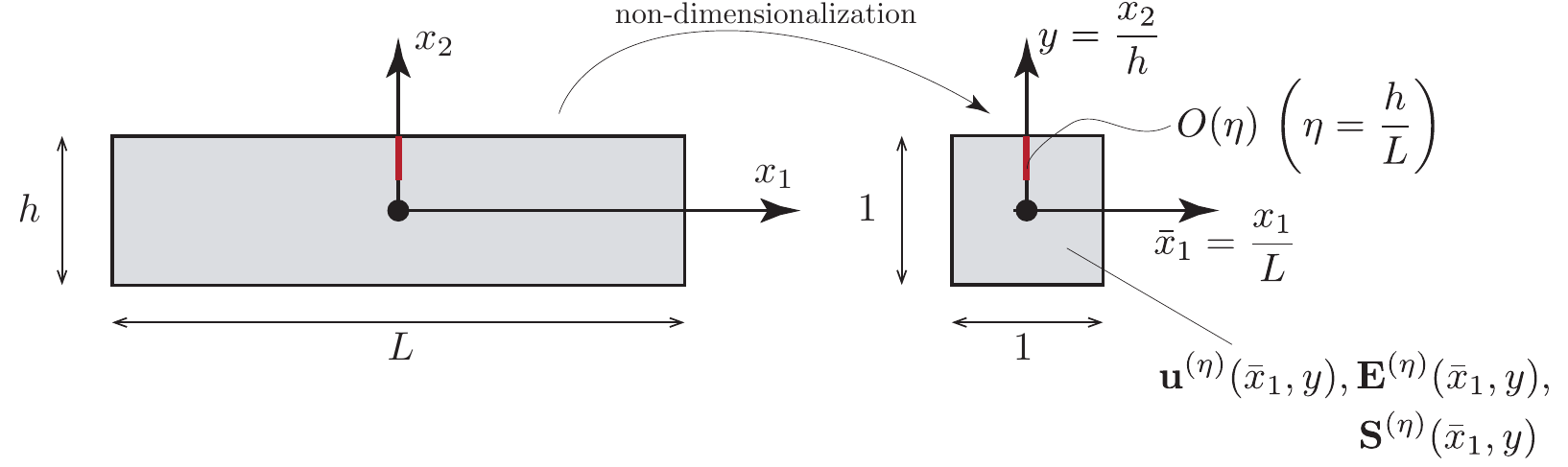}
	\caption{Geometry in the non-dimensionalized setting.}
	\label{fig:geometria1}
\end{figure}

We find it convenient to work with adimensional quantities. For this reason, we introduce the following parameters:
\begin{equation}
\bar{x}_1=\frac{x_1}{L}, \quad y=\frac{x_2}{h}, \quad \eta=\frac{h}{L}.
\end{equation}
 The domain  is then mapped to the square \([-1/2,1/2]\times[-1/2,1/2]\), see Fig. \ref{fig:geometria1}.
We  assume that  \(h \ll L \ll b\), so that $\eta\ll 1$ plays the role of a smallness parameter. We also set that the depth of the crack is of order \(\eta\) and perform a formal scalar expansion in terms of $\eta$.

 The non-dimensional displacement
is defined as
\begin{equation}
   \ub^{(\eta)}(\bar{x}_1,y)= \frac{ \ub(\xb) }{L}.
    \label{eq:non_dimensional_displacement}
\end{equation}
Here and henceforth the superscript $(\eta)$ is attached to non-dimensional quantities, intended to be expanded in terms of the smallness parameter.

Accordingly, we define the rescaled strain and stresses $\Eb^{(\eta)}$ and $\Sb^{(\eta)}$:
\begin{equation}
\Eb^{(\eta)}(\bar{x}_1, y)=\frac{1}{\eta} \, \Eb(\xb) ,\qquad
 \Sb^{(\eta)}(\bar{x}_1, y)=\frac{1}{\eta Y}\,   \Sb(\xb) ,
    \label{eq:non_dimensional_strain_stress}
\end{equation}
where   \(Y\)  is  the Young modulus of the material. For the non-dimensional bending moment and
traction we introduce the following fields:
\begin{equation}
    \begin{aligned}
         M^{(\eta)}(\bar{x}_1, y)&=\frac{1}{\eta^3 b Y L^2 }\, M(\xb) ,\\
  N^{(\eta)}(\bar{x}_1, y)&=\frac{1}{\eta^2 b Y L}\,   N(\xb) ,
    \end{aligned}
    \label{eq:non_dimensional_moments_and_tractions}
\end{equation}
which respect the assumption (\ref{eq:dimensional_scaling_assumption}).
Substituting (\ref{eq:non_dimensional_strain_stress},
\ref{eq:non_dimensional_moments_and_tractions}), into
(\ref{eq:bc_stress_distribution_dimensional}), we have that 
\begin{equation}
    S^{(\eta)}_{11} = N^{(\eta)} - 12M^{(\eta)} y,
    \label{eq:stress_distribution_non_dimensional}
\end{equation}
 and  that the following equalities hold:
\begin{equation}
    \begin{aligned}
M^{(\eta)}(\bar{x}_1, y) &= - \int_{-1/2}^{1/2} y
            S_{11}^{(\eta)}(\bar{x}_1, y) \dd y, \\
 N^{(\eta)}(\bar{x}_1, y) &= \int_{-1/2}^{1/2}  S_{11}^{(\eta)}(\bar{x}_1, y) \dd y.
    \end{aligned}
    \label{eq:bending_moments_and_traction_stresses_non_dimensional}
\end{equation}
We also introduce the shear force
\begin{equation}
    T^{(\eta)} = \frac{1}{\eta} \int_{-1/2}^{1/2} 
        S_{12}^{(\eta)}(\bar{x}_1, y) \dd y.
    \label{eq:shear_force_non_dimensional}
\end{equation}

A formal integration of the non-dimensionalized balance equation (see Appendix \ref{nondimeq} for details, and \eqref{eq:stresses_non_dimensional} in particular) through the height yields the standard balance equations for beams (in absence of loads on the mantle):

\begin{equation}\label{eq:beam_equilibrium_non_dimensional}
N^{(\eta)}_{,1} = 0, \qquad T^{(\eta)}_{,1} = 0, \qquad M^{(\eta)}_{,11} = 0.
\end{equation} 
at each order \(\eta\).

These have to be supplemented with non-dimensionalized boundary condition (see Appendix \ref{nondimeq}) in the case when no discontinuity is
present:
\begin{equation}
    \int_{-1/2}^{1/2} S_{11}^{(\eta)}\dd y = N^{(\eta)}.
    \label{eq:bc3_non_dimensional_simplified}
\end{equation}

\hypertarget{outer-asymptotic-expansion}{%
\subsection{Outer asymptotic
expansion}\label{outer-asymptotic-expansion}}
Our coarsening process is based on two asymptotic expansions: one is assumed to be valid far from the boundaries of the crack (\textit{outer expansion}); the other is instead valid in proximity of the crack, where a boundary layer is present (\textit{inner expansion}). Imposing that this inner expansion and the outer one coincide in a range where both are supposed to be valid, will yield the
matching conditions needed to close the problem and determine the terms
of the expansion. 

We will decompose the displacement field in an axial part, denoted by $w$,  and a transversal one, denoted by $v$: 
	$$
\ub=w\,\eb_1+v\,\eb_2.
$$
In order to distinguish between outer and inner expansions, we will use the same letters for the relevant fields, but different fonts. In particular, the two following sets 
	$$
\{\Sb, \Eb, \ub,  S_{\alpha\beta}, E_{\alpha\beta},\}
\qquad \{\Ssb, \Esb, \usb, \mathsf{S}_{\alpha\beta}, \mathsf{E}_{\alpha\beta}\}.
$$
denote the stress tensor, the strain tensor, the displacement field and their components, in the outer and inner expansions, respectively.

We suppose that, far from the boundaries of the crack in the middle of
the domain, the displacements, stresses and strains admit the following
asymptotic expansion in the small parameter \(\eta\):

\begin{equation}
\begin{aligned}
    \ub^{(\eta)} &= \ub^{(0)}(\bar{x}_1, y) + \eta \ub^{(1)}(\bar{x}_1, y) + \eta^2 \ub^{(2)}(\bar{x}_1, y) + \eta^3 \ub^{(3)}(\bar{x}_1, y) + o(\eta^3), \\
    \Sb^{(\eta)} &= \eta^{-2} \Sb^{(-2)}(\bar{x}_1, y) + \eta^{-1} \Sb^{(-1)}(\bar{x}_1, y) + \Sb^{(0)}(\bar{x}_1, y) + \eta \Sb^{(1)}(\bar{x}_1, y) + o(\eta), \\
    \Eb^{(\eta)} &= \eta^{-2} \Eb^{(-2)}(\bar{x}_1, y) + \eta^{-1} \Eb^{(-1)}(\bar{x}_1, y) + \Eb^{(0)}(\bar{x}_1, y) + \eta \Eb^{(1)}(\bar{x}_1, y) + o(\eta).
\end{aligned}
\label{eq:outer_asymptotic_expansion}
\end{equation}

Then, by
(\ref{eq:bending_moments_and_traction_stresses_non_dimensional}), the
asymptotic expansion of the bending moments and traction starts at order
\(-2\), and that of the shear force at order \(-3\):

\begin{equation}
    \begin{aligned}
        M^{(\eta)} &= \eta^{-2} M^{(-2)}(\bar{x}_1, y) + \eta^{-1} M^{(-1)}(\bar{x}_1, y) + M^{(0)}(\bar{x}_1, y) + \eta M^{(1)}(\bar{x}_1, y) + o(\eta), \\
        N^{(\eta)} &= \eta^{-2} N^{(-2)}(\bar{x}_1, y) + \eta^{-1} N^{(-1)}(\bar{x}_1, y) + N^{(0)}(\bar{x}_1, y) + \eta N^{(1)}(\bar{x}_1, y) + o(\eta), \\
        T^{(\eta)} &= \eta^{-3} T^{(-3)}(\bar{x}_1, y) + \eta^{-2} T^{(-2)}(\bar{x}_1, y) +\eta^{-1} T^{(-1)}(\bar{x}_1, y) + T^{(0)}(\bar{x}_1, y) + o(1).
    \end{aligned}
    \label{eq:bending_moments_asymptotic_expansion}
\end{equation}

It remains to choose the order at which the external forces act. The
relative scaling between tractions and moments was introduced previously in
(\ref{eq:dimensional_scaling_assumption}). We further assume that
\textit{the traction and bending moments will both scale so as to
generate stresses at order \(0\)}. Therefore, we get:
\begin{equation}
    \begin{aligned}
        \int_{-1/2}^{1/2} S^{(0)}_{11}\text{d}y &= N^{(0)} = N^{(\eta)}_o,\\
        -\int_{-1/2}^{1/2} y S^{(0)}_{11}\text{d}y &= M^{(0)} = M^{(\eta)}_o,
    \end{aligned}
    \label{eq:bc3_non_dimensional_simplified_ext_forces}
\end{equation}
(see Eqs.
\eqref{eq:bc3_non_dimensional}, \eqref{eq:bc4_non_dimensional} in the Appendix \ref{nondimeq}).

We now present some relevant solutions of the non-dimensional problems at each order, referring to the Appendix \ref{outerappendix} for the details. 

At the order $-2$, we get that the stress and the strain are null $	\Sb^{(-2)}=\Eb^{(-2)} = \bm{0} $, while 
$ \ub^{(0)}=\ub^{(0)}(\bar{x}_1)$.

At the order $-1$, the stress and the strain are still null $\Sb^{(-1)}=\Eb^{(-1)} = \bm{0} $, while the following expression is found for the displacement field:
\begin{equation}\label{displ-1}
 \ub^{(1)} =\left( w^{(1)}(\bar{x}_1) - y v^{(0)}_{,1}(\bar{x}_1)\right) \eb_1 +
	v^{(1)}(\bar{x}_1) \eb_2.
\end{equation}
We here recognize the  Anstatz customarily adopted for the standard Euler-Bernoulli displacement field.

At the order 0 and 1 all the fields are non-null (see Appendix \ref{outerappendix}); in particular, the displacement turns out to be

\begin{multline}\label{displ0}
\ub^{(2)}=\left(   w^{(2)}(\bar{x}_1) -y v^{(1)}_{,1}(\bar{x}_1) \right)\eb_1+\\
\left(  v^{(2)}(\bar{x}_1) -\frac{\nu}{1-\nu}
\left(w^{(1)}_{,1}(\bar{x}_1)y - v^{(0)}_{,11}(\bar{x}_1) \frac{y^2}{2}\right) \right)\eb_2.
\end{multline}
On comparing \eqref{displ0} and \eqref{displ-1}, we notice that in the transversal component two corrective terms appear: the former depends on the axial strain $w^{(1)}_{,1}$  and is linear in the transversal coordinate $y$; the latter depends on the curvature $v^{(0)}_{,11}$ and is quadratic in $y$. The influence of both rely upon the material, through the Poisson coefficient $\nu$.

The expansion up to order $1$ finally allows to determine the constitutive equation for the bending moment and the axial force, once 
%
%
\eqref{eq:bending_moments_and_traction_stresses_non_dimensional} is used:
\begin{equation}
    \begin{aligned}
        M^{(0)}  = \frac{1}{12(1-\nu^2)}\, v^{(0)}_{,11},\\
        M^{(1)}  = \frac{1}{12(1-\nu^2)}\, v^{(1)}_{,11},\\
    \end{aligned}
    \label{eq:bending_moments_order_0_1}
\end{equation}
\begin{equation}
    \begin{aligned}
        N^{(0)} = \frac{1}{1-\nu^2}\, w^{(1)}_{,1},\\
        N^{(1)} = \frac{1}{1-\nu^2}\,w^{(2)}_{,1}.\\
    \end{aligned}
    \label{eq:tractions_order_0_1}
\end{equation}

\hypertarget{inner-expansion}{%
\subsection{Inner asymptotic expansion}\label{inner-expansion}}

\begin{figure}
	\centering
	\includegraphics[scale=0.9]{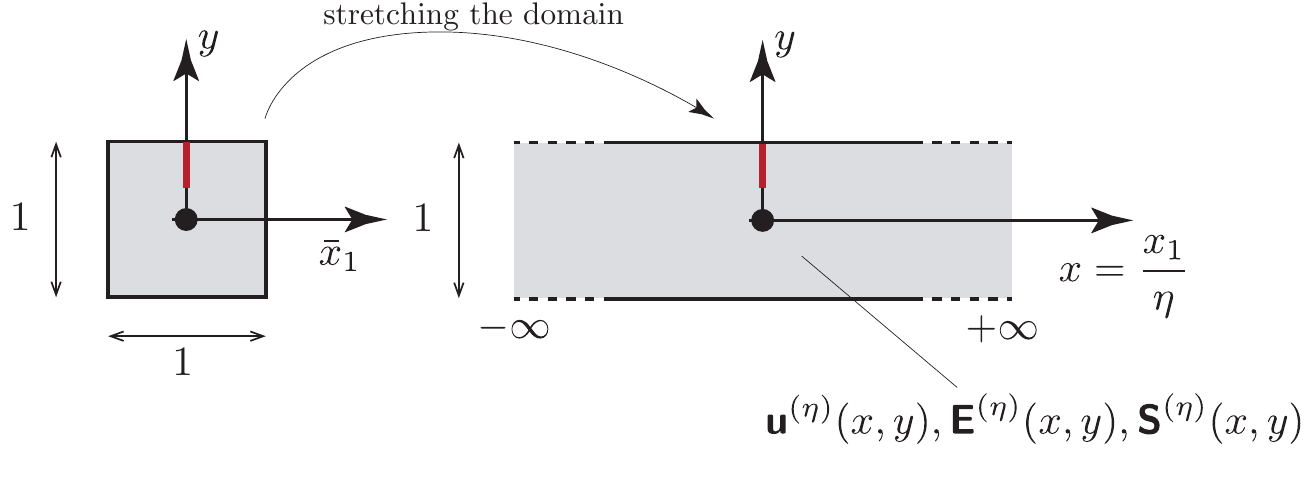}
	\caption{Stretched domain instrumental to assess the matching conditions.}
	\label{fig:geometria2}
\end{figure}

The asymptotic expansions (\ref{eq:outer_asymptotic_expansion}) are not
valid in proximity of the crack, since a boundary layer is present.
Another inner asymptotic expansion has to be derived, after a rescaling of the
coordinates.  We only consider a crack with rectangular
shape, of vanishing thickness, symmetrical with respect to the \(y\)
axis (each side of the line of defect is a line parallel to the \(y\)
axis). This crack will originate from the upper boundary of the domain,
that is from \(y = \frac{1}{2}\). More specifically, we introduce the
rescaled variable

\begin{equation}
    x = \frac{\bar{x}_1}{\eta},
\end{equation}
and the inner expansion:
\begin{equation}
\begin{aligned}
    \ub^{(\eta)} &= \usb^{(0)}(x, y) + \eta \usb^{(1)}(x, y) + \eta^2 \usb^2(x, y) + \eta^3 \usb^3(x, y) + \cdots, \\
    \Sb^{(\eta)} &= \eta^{-2} \Ssb^{(-2)}(x, y) + \eta^{-1} \Ssb^{(-1)}(x, y) + \Ssb^{(0)}(x, y) + \eta \Ssb^{(1)}(x, y) + \cdots, \\
    \Eb^{(\eta)} &= \eta^{-2} \Esb^{(-2)}(x, y) + \eta^{-1} \Esb^{(-1)}(x, y) + \Esb^{(0)}(x, y) + \eta \Esb^{(1)}(x, y) + \cdots
\end{aligned}
\label{eq:inner_asymp_expansion}
\end{equation}
(see Fig. \ref{fig:geometria2})

The functions \(\usb^{(i)}, \Esb^{(i)}, \Ssb^{(i)}\) are defined in
the perturbed domain, which is infinitely extended in the \(x\)
direction, in contrast with the functions of the outer expansion, which
are defined on the strip
\((\bar{x}_1, y) \in [-1/2, 1/2] \times [-1/2, 1/2]\). To obtain the
matching conditions, starting from the displacements, the general
expression is

\begin{equation}
    \ub^{(\eta)} = \sum_i \eta^i \ub^{(i)}(\eta x, y).
    \label{eq:u_rescaled}
\end{equation}

Each term is expanded in series with respect to \(\bar{x}_1\) near the
crack, \(\bar{x}_1 = 0\):

\begin{equation}
    \ub^{(i)}(\eta x, y) = \ub^{(i)}(0, y) + \eta x \ub^{(i)}_{,1}(0, y) + \frac{1}{2} \eta^2 x^2 \ub^{(i)}_{,11}(0, y) + o(\eta^2).
    \label{eq:u_rescaled_taylor}
\end{equation}

Inserting (\ref{eq:u_rescaled_taylor}) into (\ref{eq:u_rescaled}), and
comparing with equation (\ref{eq:inner_asymp_expansion}), yields the
matching conditions between inner and outer expansion:

\begin{equation}
\begin{aligned}
    \lim_{x\to\pm\infty} \usb^{(0)}(x, y) &= \ub^{(0)}(0^\pm, y),\\
    \lim_{x\to\pm\infty} \usb^{(1)}(x, y) &= \ub^{(1)}(0^\pm, y) 
        + \lim_{x\to\pm\infty} x\ub^{(0)}_{,1}(0^\pm, y),\\
    \lim_{x\to\pm\infty} \usb^{(2)}(x, y) &= \ub^{(2)}(0^\pm, y) 
        + \lim_{x\to\pm\infty} \left(x\ub^{(1)}_{,1}(0^\pm, y) +
        \frac{1}{2} x^2 \ub^{(0)}_{,11}(0^\pm, y)\right),\\
\end{aligned}
\label{eq:u_matching_conditions}
\end{equation}

Here, the functions are evaluated at \(0^\pm\), to indicate that the
results of the two limits, \(x\to\pm\infty\), could be different, and
thus the functions discontinuous, at \(\bar{x}_1 = 0\). The matching
conditions for the stresses and strains are of the same kind, except
that the expansions have to start at order \(-2\) instead of \(0\).

The solution process starts again at order \(-2\), where the boundary
conditions for \(\lim_{x\to\pm\infty}\), are given by the matching
conditions. These equations are defined in the rescaled domain
\(\Omega\), which is extended to the interval \(x \to -\infty\) to \(x \to \infty\), and
from \(y = -\frac{1}{2}\) to \(y = \frac{1}{2}\) except in the region of
the crack, which has a non-dimensional depth \(a\). In
the following we will denote with \(\partial \Omega^+\) the part of the
boundary of \(\Omega\) excluding the bases at \(x \to \pm\infty\), that
is, the part of the boundary which is always stress free. We here report the relevant results, and refer to Appendix \ref{innerappendix} for the details. 

At the order $-2$ , both the strain and the stress are null, 
\(\Esb^{(-2)}(\usb^{(0)}) = \Ssb^{(-2)}=\mathbf{0}\),  and then 
 the field \(\usb^{(0)}\) is a rigid displacement:
\begin{equation}
	\usb^{(0)}(x, y) = \tb^{(0)} + \omega^{(0)}\eb_3 \times \yb,
\end{equation}
where we introduced the vector \(\yb =x\eb_1+y\eb_2\), and \(\tb^{(0)}\),
\(\omega^{(0)}\), both constant, are a translation and a rotation, respectively. Since from the outer
expansion we know that \(\ub^{(0)} = v^{(0)}(\bar{x}_1)\eb_2\), it follows
that the \(\eb_1\) component must
vanish
\begin{equation}
	t^{(0)}_1 - \omega^{(0)} y  = 0 \qquad \forall y,
\end{equation}
which in turn implies \(\omega^{(0)} = 0 = t^{(0)}_1\), and
\(v^{(0)}(0^\pm) = t^{(0)}_2\), from which the continuity of \(v^{(0)}\) at \(0\)
follows:

\begin{equation}
	[\![ v^{(0)} ]\!] = 0 \qquad \text{at } \bar{x}_1 = 0,
\end{equation}
where the jump at a specific value of \(\bar{x}_1\) is defined as
\([\![ f ]\!] = f(\bar{x}_1^+)-f(\bar{x}_1^-)\).

At the order $-1$, we get again that the strain and the stress are null
\(\Esb^{(-1)}(\usb^{(1)}) = \mathbf{0}= \Ssb^{(-1)}\), which
implies that the displacement field at order \(-1\) is also rigid. Now the matching conditions imply that 
\begin{equation}
	[\![ v^{(1)} ]\!] = [\![ v^{(0)}_{,1} ]\!] = [\![ w^{(1)} ]\!] = 0
	\qquad \text{at } \bar{x}_1 = 0.
\end{equation}
and the expression of \(\usb^{(1)}\) is determined:
\begin{equation}
	\usb^{(1)}(x, y) =  \left(w^{(1)}(0) - y v^{(0)}_{,1} (0) \right) \eb_1 + 
	\left(x v^{(0)}_{,1} (0) + v^{(1)}(0) \right) \eb_2.
\end{equation}

At the order 0, the problem to be solved is
\begin{equation}
	\begin{cases}
		&   \div\Ssb^{(0)} = \bm{0},\\
		&  \Esb^{(0)}  = -\nu \text{tr}\Ssb^{(0)} \Ib
		+ (1 + \nu) \Ssb^{(0)},\\
		&  \Ssb^{(0)} \nb = \bm{0} \hfill \text{on } \partial \Omega^+,\\
		&   \displaystyle\lim_{x\to\pm\infty} \Ssb^{(0)}(x, y) = \Sb^{(0)}(0^\pm, y), 
	\end{cases}
\end{equation}
where the last equation  comes from the matching conditions.  We get the  following jump conditions:  

\begin{equation}
	\begin{aligned}
	&[\![ v^{(0)}_{,11} ]\!] =[\![ M^{(0)} ]\!] = 0,\\
&[\![ w^{(1)}_{,1} ]\!]=[\![N^{(0)}]\!] =  0 \qquad \text{at } \bar{x}_1 = 0.
	\end{aligned}
\label{jumpord0}
\end{equation}

These  results imply that the stress \(S^{(0)}_{11}\) is continuous
at \(\bar{x}_1 = 0\). We cannot solve the first equation of
\eqref{jumpord0} by choosing directly the value the
stresses have to satisfy at infinity, given by the outer solution at
order \(0\):
\begin{equation}
        \mathsf{S}^{(0)}_{22} = 0 = \mathsf{S}^{(0)}_{12}, \quad \mathsf{S}^{(0)}_{11} = S^{(0)}_{11}(0) = 
            \frac{w^{(1)}_{,1}(0)-yv^{(0)}_{,11}(0)}{1-\nu^2},\\
    \label{eq:gamma_sigma_inner_0}
\end{equation}
since this stress tensor does not satisfy the condition that, on the
boundary of the crack, \(\Ssb^{(0)} \nb  = \mathbf{0}\).
This will result in a \textit{boundary layer}. 

We introduce two pairs of
auxiliary stress, strain and displacement fields, (\(\bm{\Sigma}^N\), \(\bm{\Gamma}^N\)),
(\(\bm{\Sigma}^M\),  \(\bm{\Gamma}^M\), \(\zetab^M\)), 
%
%
defined in order to satisfy the
following auxiliary problems:
\begin{equation}
\begin{cases}
       & \div\bm{\Sigma}^N = \bm{0},   \hfill \text{in } \Omega, \\
       & \bm\Gamma(\zetab^N) = -\nu \text{tr}\bm{\Sigma}^N \Ib
            + (1 + \nu) \bm{\Sigma}^N, \hfill \text{in } \Omega, \\
      &  \bm{\Sigma}^N \nb = -\frac{1}{1-\nu^2}
            \mathbf{e}_1 \hfill \text{on } \partial \Omega^+, \\
        &\displaystyle\lim_{x\to\pm\infty} \bm{\Sigma}^N = \bm{0}.
        \end{cases}
    \label{eq:aux_inner_problem_order_0_traction}
\end{equation}

\begin{equation}
    \begin{cases}
        &\div\bm{\Sigma}^M = \bm{0},   \hfill \text{in } \Omega, \\
        &\bm\Gamma(\zetab^M)  = -\nu \text{tr}\bm{\Sigma}^M \Ib
            + (1 + \nu) \bm{\Sigma}^M, \hfill \text{in } \Omega, \\
        &\bm{\Sigma}^M \nb = \frac{y}{1-\nu^2}
            \mathbf{e}_1 \hfill \text{on } \partial \Omega^+, \\
        &\displaystyle\lim_{x\to\pm\infty} \bm{\Sigma}^M = \bm{0}.
    \end{cases}
    \label{eq:aux_inner_problem_order_0_bending}
\end{equation}

\begin{figure}
	\centering
	\includegraphics[scale=.8]{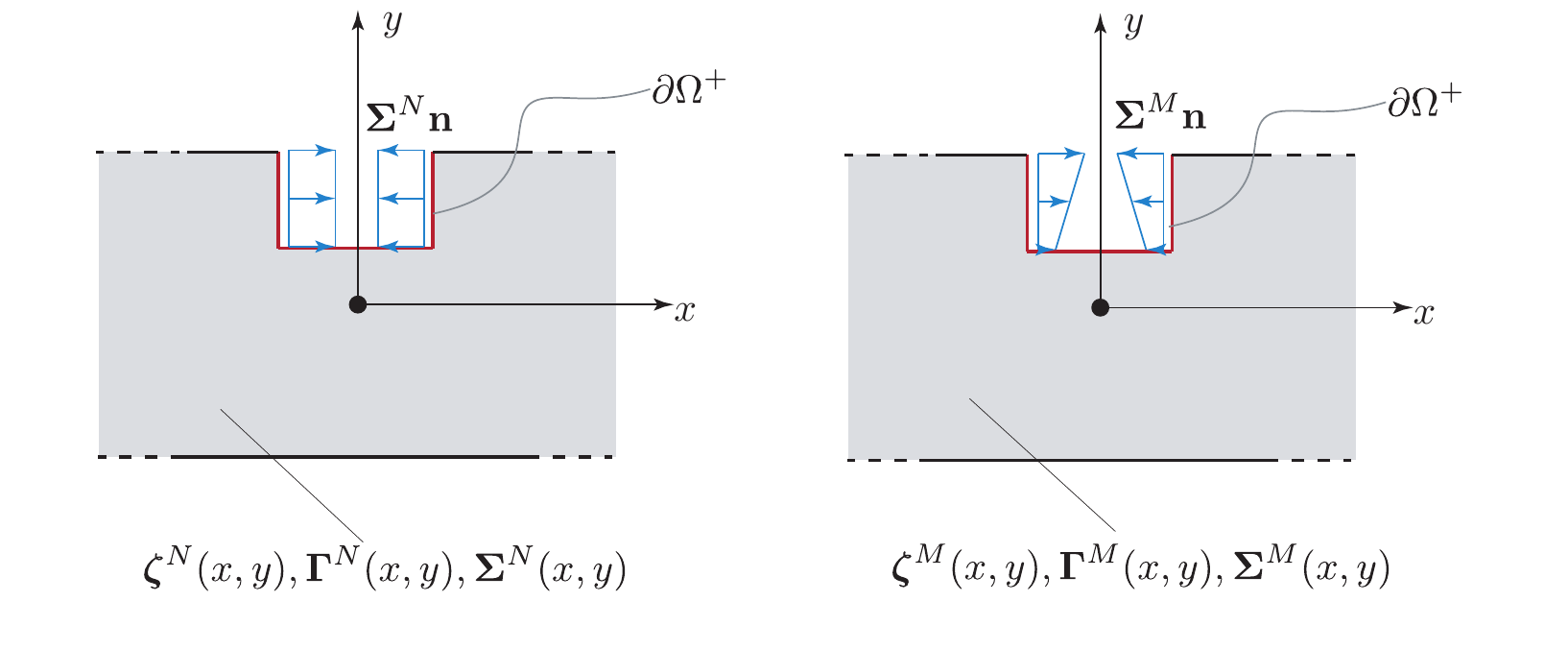}
	\caption{Auxiliary stress fields $\bm\Sigma^N$ and $\bm\Sigma^M$. For sake of clarity, in the figure the crack is stretched in direction of $x$, but the two lips have to be intended as in contact. }
	\label{fig:auxiliary}
\end{figure}

By linearity, the displacement will be the sum of three components:
 \(\zetab^C\), which
satisfies the condition at infinity (yielding the outer solution), 
\(\zetab^N\) and \(\zetab^M\), vanishing far from the crack,  such
that the boundary conditions of the two inner problems be satisfied.

Imposing that
\begin{equation}
\begin{aligned}
&\Gamma_{11}(\zetab^C)=E_{11}^{(0)}= w^{(1)}_{,1}(0) - y v^{(0)}_{,11}(0),\\
&\Gamma_{22}(\zetab^C)=E_{22}^{(0)}=  \frac{-\nu}{1-\nu}\left(w^{(1)}_{,1}(0) -y v^{(0)}_{,11}(0)\right),
\end{aligned}
\end{equation}
%
%
%
gives the expression of \(\zetab^C\):
\begin{equation}
    \zetab^C =\left( w^{(1)}_{,1}(0)x - x y v^{(0)}_{,11}(0)\right)\eb_1+        \left(-\frac{\nu}{1-\nu}w^{(1)}_{,1}(0)y +
            \left(\frac{\nu y^2}{2(1-\nu)} + \frac{x^2}{2}\right) v^{(0)}_{,11}(0)\right)\eb_2
\end{equation}

Thus, the general expression of the solution, determined up to a rigid
motion, is given by:

\begin{equation}
    \begin{aligned}
        w^{(2)} &= t^{(2)}_1 - \omega^{(2)} y + w^{(1)}_{,1}x - x y v^{(0)}_{,11}
            +w^{(1)}_{,1}\zeta^N_1+ v^{(0)}_{,11}\zeta^M_1, \\
        v^{(2)} &=  t^{(2)}_2 +\omega^{(2)} x -\frac{\nu}{1-\nu}w^{(1)}_{,1}y +
            \left(\frac{\nu y^2}{2(1-\nu)} + \frac{x^2}{2}\right) v^{(0)}_{,11}
            + w^{(1)}_{,1}t\zeta^N_2 + v^{(0)}_{,11}\zeta^M_2,
    \end{aligned}
    \label{eq:solution_general_inner_0_v}
\end{equation}
where, for sake of brevity, it is implicit that \(w^{(1)}_{,1}\) and
\(v^{(0)}_{,11}\) are evaluated at \(\bar{x}_1 = 0\), where both functions
are continuous, and that \(\zetab^N\) and \(\zetab^M\) both are
functions of \(x, y\). 


The expression of the components of \(\zetab^N\) at infinity can be
obtained by noticing that, if \(\bm{\Sigma}^N\) tends to zero, so does
\(\bm\Gamma^N(\zetab^N)\), due to the constitutive relations. Therefore,
the field \(\zetab^N\) will be rigid at infinity. The same is true for
\(\zetab^M\):

\begin{equation}
    \begin{aligned}
        \lim_{x\to+\infty}\left( \zetab^N - (\Phi^N)^+ \eb_3\times\yb\right) 
            &= (H_w^N)^+\eb_1+(H_v^N)^+\eb_2, \\
        \lim_{x\to-\infty}\left( \zetab^N - (\Phi^N)^- \eb_3\times\yb \right) =
            &(H_w^N)^-\eb_1+(H_v^N)^-\eb_2,
    \end{aligned}
    \label{eq:v_t_at_infinity_inner_0}
\end{equation}

\begin{equation}
    \begin{aligned}
        \lim_{x\to+\infty}\left( \zetab^M - (\Phi^M)^+ \eb_3\times\yb\right) 
            &= (H^M_w)^+\eb_1+(H^M_v)^+\eb_2, \\
        \lim_{x\to-\infty}\left( \zetab^M - (\Phi^M)^- \eb_3\times\yb \right) =
            &= (H^M_w)^-\eb_1+(H^M_v)^-\eb_2,
    \end{aligned}
    \label{eq:v_b_at_infinity_inner_0}
\end{equation}
and we define the respective differences
\begin{equation}
    \begin{aligned}
        \Phi^N &= (\Phi^N)^+ - (\Phi^N)^-, \quad  H_w^N=(H_w^N)^+-(H_w^N)^-  \quad H_v^N=(H_v^N)^+-(H_v^N)^-\\
        \Phi^M &= (\Phi^M)^+ - (\Phi^M)^-, \quad  H^M_w=(H^M_w)^+-(H^M_w)^-  \quad H^M_v=(H^M_v)^+-(H^M_v)^-
    \end{aligned}
    \label{eq:diff_Cd_Kd}
\end{equation}
$ \Phi^I$, $ H^I_j$ with $I=(N,M)$ and $j=(w,v)$ then represent \textit{the relative rotation, axial ($j=w$) and transversal ($j=v$) displacement between the bases when the stress $\Sigmab^I\nb$ is applied on the crack}.

Now, it is necessary to (i) substitute  into the third equation of
(\ref{eq:u_matching_conditions}), these previous results and the
expressions of \(\ub\) up to second order; (ii)  gather the terms with the same scaling and recall equations
(\ref{eq:v_t_at_infinity_inner_0}, \ref{eq:v_b_at_infinity_inner_0}). After some manipulations (see Appendix \ref{innerappendix}), we get: 
\begin{equation}
	\begin{aligned}
		{}[\![v^{(1)}_{,1}]\!] &= \Phi^N(1-\nu^2)N^{(0)}(0) + 12 \Phi^M (1-\nu^2)M^{(0)}(0),\\
		{}[\![v^{(2)}]\!] &= H_v^N(1-\nu^2)N^{(0)}(0) +12H^M_v (1-\nu^2) M^{(0)}(0),\\
		{}[\![w^{(2)}]\!] &= H_w^N(1-\nu^2)N^{(0)}(0) + 12H^M_w(1-\nu^2)M^{(0)}(0).
	\end{aligned}
	\label{eq:alternative_transmission_conditions_order_0}
\end{equation}
The constants $\Phi^I$, $H^I_j$ ($I=(N,M)$ and $j=(w,v)$) are then \textit{compliance coefficients}, which depend on the depth of the crack.

\section{A constitutive model of a cracked beam}
\subsection{The compliance coefficients}\label{appendix-a-calculation-of-the-coefficients}

The physical meaning of the compliance coefficients $ \Phi^I$, $ H^I_j$ as relative rotations and displacements suggests a possible way to compute them, by the use of the virtual work principle. In particular, the relative rotation between the bases can be determined introducing an auxiliary problem, where a unit torque is applied at the bases. Let  $\Tb^M$  denote the stress tensor solving this problem. It can be determined by superimposing the stress field solution of the outer problem $\Sb^{(0)}$ evaluated at $M=1$ and the boundary layer stress:
\begin{equation}
\Tb^M=\Sb^{(0)}|_{M=1} + 12(1-\nu^2)\bm{\Sigma}^M.\\
\label{TM}
\end{equation}

\begin{figure}
	\centering
	\includegraphics[width=0.8\linewidth]{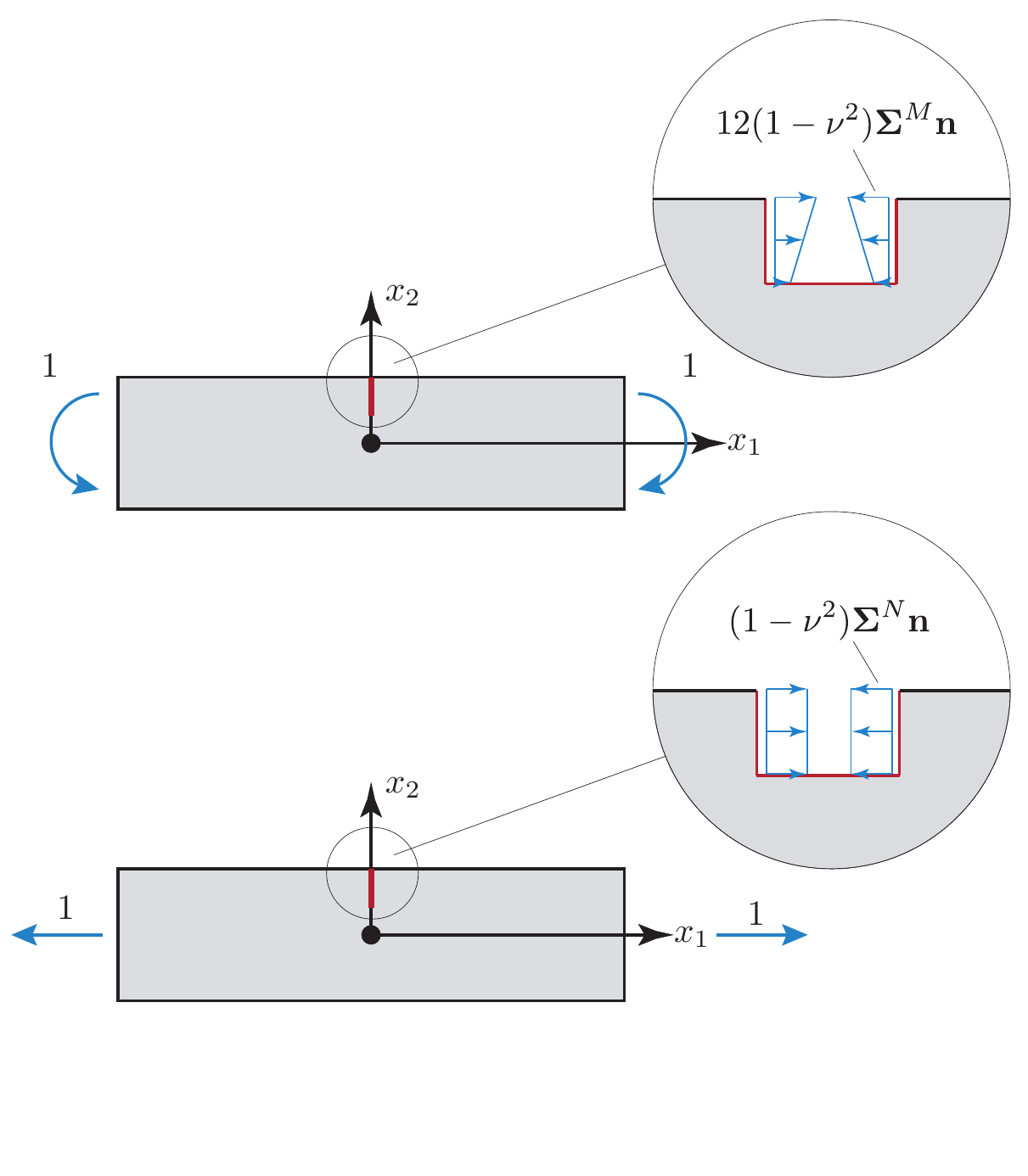}
	\caption{Auxiliary problems instrumental to compute the compliance coefficients.}
	\label{fig:auxiliarym}
\end{figure}
By the principle of virtual works, the  work of the external forces of this system evaluated over the displacements of the real system has to be equal to the the internal work, given by the integral over the all domain of the scalar product between the stress $\bm{T}^{M}$ 
 of the auxiliary system and the strain of the real system $\mathbbm{C}^{-1}\bm{\Sigma}^N$, where \(\mathbbm{C}^{-1}\) is the fourth order  compliance tensor, such that 
 \(\Esb =\mathbbm{C}^{-1}\Ssb= (1+\nu)\bm{\Sigma} - \nu\text{tr}(\bm{\Sigma})\Ib\). Therefore, we get
\begin{equation}
	1\cdot\big((\Phi^N)^+-(\Phi^N)^-\big) =\Phi^N = \int_{\Omega} \bm{T}^{M} \cdot
\mathbbm{C}^{-1}\bm{\Sigma}^N \text{d}x \text{d}y.\\
\end{equation}
Analogous expressions can be determined for the others coefficients:
\begin{equation}
	\begin{aligned}
		\Phi^M &= 1\cdot\big((\Phi^M)^+-(\Phi^M)^-\big) = \int_{\Omega} \bm{T}^{M} \cdot
		\mathbbm{C}^{-1}\bm{\Sigma}^M \text{d}x \text{d}y, \\
		H_w^N &= 1\cdot\big((H_w^N)^+-(H_w^N)^-\big)  = \int_{\Omega} \bm{T}^{N} \cdot
		\mathbbm{C}^{-1}\bm{\Sigma}^N \text{d}x \text{d}y, \\
		H^M_w &= 1\cdot\big((H^M_w)^+-(H^M_w)^-\big)= \int_{\Omega} \bm{T}^{N} \cdot
		\mathbbm{C}^{-1}\bm{\Sigma}^M \text{d}x \text{d}y, \\
	\end{aligned}
\label{compliance}
\end{equation}
where 
\begin{equation}
	\begin{aligned}
		\bm{T}^{N} &= \Sb^{(0)}|_{N=1} + (1-\nu^2)\bm{\Sigma}^N, 
		\label{TN}
	\end{aligned}
\end{equation}
where \(\Sb^{(0)}|_{N=1}\) 
is  the stress fields solution to the outer
problem such that the axial forces at the bases are
\(N=1\).

\begin{figure}
	\centering
	\includegraphics[scale=.8]{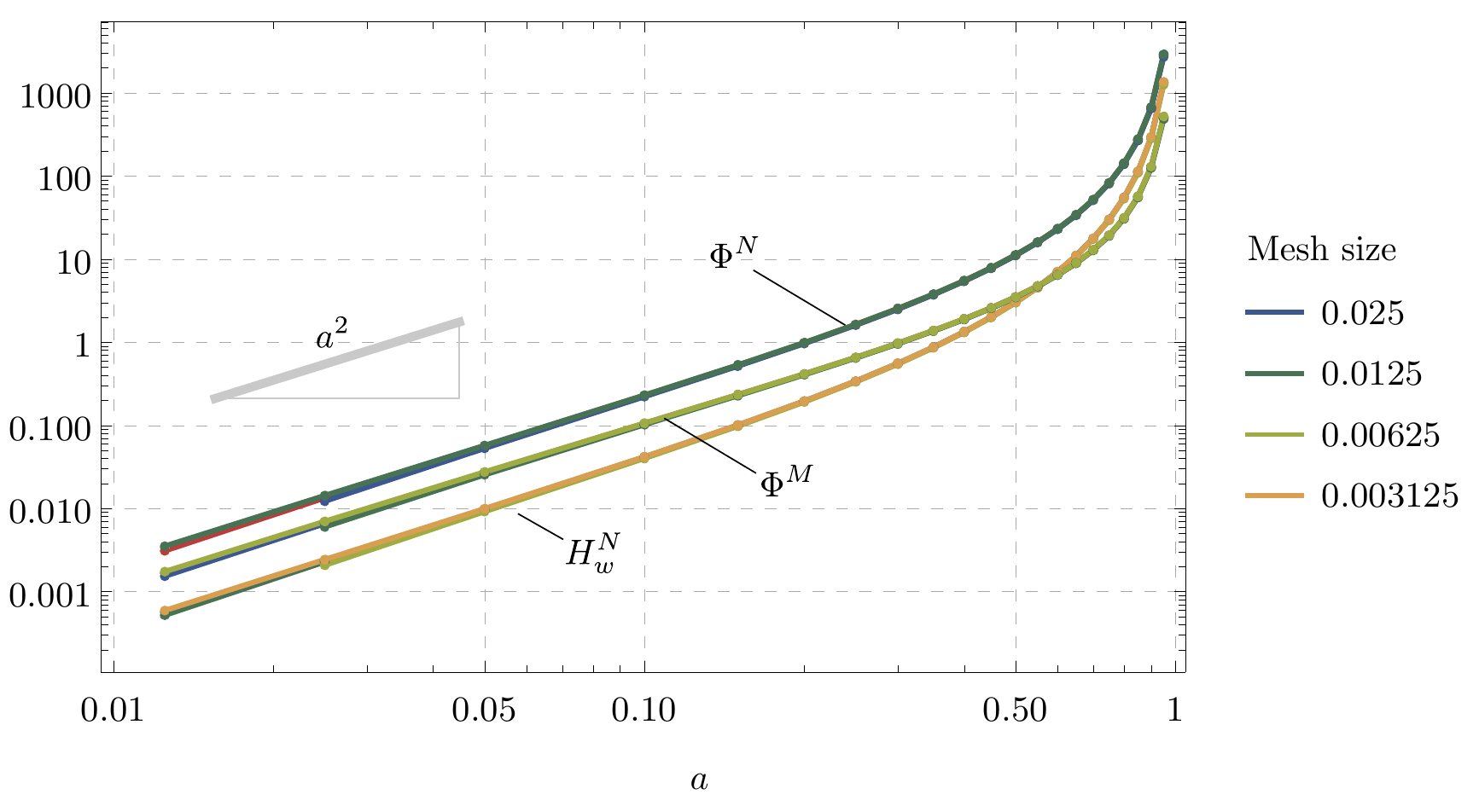}
	\caption{Convergence of the coefficients $H_w^N(a)$, $\Phi^M(a)$ and $\Phi^N(a)$ in logarithmic scale, for decreasing mesh size in the region surrounding the crack. }
	\label{fig:conv_Cb}
\end{figure}

It is therefore necessary to  solve the auxiliary problems
(\ref{eq:aux_inner_problem_order_0_traction},
\ref{eq:aux_inner_problem_order_0_bending}), and determine the stress
fields \(\bm{\Sigma}^N\) and \(\bm{\Sigma}^M\). We solve these problems
numerically by means of a finite element discretisation. Thanks to
symmetry, half of the domain, that is the rectangle
\([0,n]\times[-1/2,1/2]\), with \(n\) a given number such that
\(n\gg 1\), is discretised. On the upper and lower boundaries,
\([0,n]\times \pm 1/2\), and on the right base, \(n \times [-1/2,1/2]\),
an homogeneous Neumann condition for the stress is assigned. At the left
boundary instead, two subdivisions are considered: on the interval
\(0\times[1/2-\alpha,1/2]\), that is the boundary of the
crack, a non-homogeneous Neumann condition is assigned, coming from the
auxiliary problem being solved. The rest of the boundary is a symmetry
plane, on which the \(x\) component of the displacement is set to zero.
The elasticity problem is discretised with
$P_1$ Lagrange elements, and solved using the
\texttt{FEniCS} finite element library.  We checked that the grid was fine
enough to yield the converged results for the coefficients. In figure
\ref{fig:conv_Cb}, we report the results for the coefficients $H_w^N$, $\Phi^M$, and $\Phi^N$  for
different mesh sizes, showing a convergent behaviour for a decreasing
mesh size. We also notice that the asymptotic  behavior of all coefficients is $O(a^2)$.

\hypertarget{beam-problem}{%
\subsection{Regularization}\label{beam-problem}}

Summarizing the previous results, the displacement \(v^{(0)}\), shear force
\(T^{(0)}\), and bending moment \(M^{(0)}\) do not depend on the
presence of the crack at \(\bar{x}_1 = 0\)

\begin{equation}
    [\![v^{(0)}]\!] = 0,\quad [\![v^{(0)}_{,1}]\!] = 0, 
    \quad [\![T^{(0)}_1]\!] = [\![M^{(0)}]\!] = [\![N^{(0)}]\!] = 0.
    \label{eq:beam_problem_order_0}
\end{equation}

On the other hand, the displacements \(v^{(1)}\) and \(w^{(1)}\), satisfy the
following transmission conditions

\begin{equation}
    \begin{aligned}
        {}[\![w^{(1)}]\!] &= [\![w^{(1)}_{,1}]\!] = [\![v^{(1)}]\!] = 0, \\
        {}[\![v^{(1)}_{,1}]\!] &= \Phi^N(1-\nu^2)N^{(0)}(0) + 12 \Phi^M (1-\nu^2)M^{(0)}(0).
    \end{aligned}
    \label{eq:beam_problem_order_1_1}
\end{equation}

At next order:

\begin{equation}
    \begin{aligned}
        {}[\![v^{(2)}]\!] &= H_v^N(1-\nu^2)N^{(0)}(0) + 12H^M_v (1-\nu^2) M^{(0)}(0),\\
        {}[\![w^{(2)}]\!] &= H_w^N(1-\nu^2)N^{(0)}(0) + 12H^M_w(1-\nu^2)M^{(0)}(0).
    \end{aligned}
    \label{eq:beam_problem_order_2}
\end{equation}

Some of these transmission conditions are non-zero, due to the presence
of a boundary layer around the boundary of the defect. We are interested
in the first order at which a correction to the energy of the cracked
beam, due do the boundary layers, appears. The values of the coefficients are found by solving
the auxiliary problems (\ref{eq:aux_inner_problem_order_0_traction},
\ref{eq:aux_inner_problem_order_0_bending}) numerically, as shown in
appendix A.


The potential energy of the beam will be minus one half the work done by
the external forces. Let us consider a small part of the beam, including
the crack region, spanning the interval \([-\bar{\ell}/2, \bar{\ell}/2]\),
with \(\bar{\ell} = \ell /L\). The discontinuity of solutions happens at
\(\bar{x}_1 = 0\), and a possible regularization is to have the jumps
linearly spread along the interval, such as
\([\![w^{(2)}]\!] \bar{x}_1/\ell\) and \([\![v^{(1)}_{,1}]\!] \bar{x}_1/\ell\).
The constitutive relation inside the interval will then be different
than that in the outer region. Considering first the case when the beam
is loaded with an external traction but no external bending moment,
\(N^{(\eta)}_o \neq 0\), \(M^{(\eta)}_o = 0\), we have, using the
transmission conditions (\ref{eq:transmssion_conditions_order_0}):

\begin{equation}
    \begin{aligned}
        w_{,1} &= \eta\left( 1+ \frac{\eta H_w^N}{\bar{\ell}} \right) w^{(1)}_{,1} 
        = \mathsf{\bar{H}_N} N^{(\eta)} = \frac{\mathsf{\bar{H}_N}}{1-\nu^2} w^{(1)}_{,1}, \\
        v_{,11} &= \frac{\eta \Phi^N}{\bar{\ell}} w^{(1)}_{,1} = \mathsf{\bar{H}_{NM}} N^{(\eta)}
        = \frac{\mathsf{\bar{H}_{NM}}}{1-\nu^2} w^{(1)}_{,1}. 
    \end{aligned}
\end{equation}

In the case \(N^{(\eta)}_o = 0\), \(M^{(\eta)}_o \neq 0\), instead

\begin{equation}
    \begin{aligned}
        w_{,1} &= \frac{\eta^2 H^M_w}{\bar{\ell}} v^{(0)}_{,11} 
        = \mathsf{\bar{H}_{NM}} M^{(\eta)} = \frac{\mathsf{\bar{H}_{NM}}}{12(1-\nu^2)} v^{(0)}_{,11}, \\
        v_{,11} &= \left(1+\frac{\eta \Phi^M}{\bar{\ell}} \right)v^{(0)}_{,11} = \mathsf{\bar{H}_{M}} M^{(\eta)}
        = \frac{\mathsf{\bar{H}_{M}}}{12(1-\nu^2)} v^{(0)}_{,11}.
    \end{aligned}
\end{equation}
By superimposing the effects, we then have 
\begin{equation}
	\begin{aligned}
&\mathsf{\bar{H}_N}(a)=(1-\nu^2)\left( 1+ \frac{\eta H_w^N(a)}{\bar{\ell}} \right) \eta,\\
&\mathsf{\bar{H}_{NM}}(a) =(1-\nu^2)\frac{\eta \Phi^N(a)}{\bar{\ell}}=12(1-\nu^2)\frac{\eta^2 H^M_w(a)}{\bar{\ell}} ,\\
&\mathsf{\bar{H}_M}(a) =12(1-\nu^2)\left(1+\frac{\eta \Phi^M(a)}{\bar{\ell}} \right),
	\end{aligned}
\end{equation}
where we have made explicit the dependence on the crack depth $a$.
That $\Phi^N=12H^M_w$ has been confirmed by numerical results reported below. 
To pass to the dimensional form, we use \eqref{eq:non_dimensional_moments_and_tractions} and get:
\begin{equation}
	\begin{aligned}
		\mathsf{H_N}(a)  &= \frac{(1-\nu^2)}{bhY}
		\Big( 1+\frac{h }{\ell}  H_w^N(a)\Big), \\
		\mathsf{H_{NM}}(a) &= \frac{(1-\nu^2)}{bh^2Y} \frac{h}{\ell}\Phi^N(a) = 
		\frac{12(1-\nu^2)}{bh^2Y} \frac{h}{\ell} H^M_w(a),\\
		\mathsf{H_M}(a)  &= \frac{12(1-\nu^2)}{bh^3Y}
		\Big.\Big( 1+\frac{h}{\ell}\Phi^M(a)\Big.\Big).
	\end{aligned}
	\label{eq:beam_stiffness_coeff_mat_dimensional}
\end{equation}

The energy of the beam, in the interval $\ell$ can be then written as
\begin{equation}
	\Pc^{\ell}(a) =\frac{1}{2} 
	\int_{-\frac{\ell}{2}}^{\frac{+\ell}{2}} \left(\mathsf{H_N}(a) N^2+2\mathsf{H_{NM}}(a) NM +\mathsf{H_{M}}(a) M^2 \right) \,\text{d} x_1.
	\label{eq:potential_energy_beam_dimensional_lc_forces}
\end{equation}

The energy of the beam can be more suitably written in terms of the stiffness coefficient, instead of the compliance one:
\begin{equation}
		\Uc^{\ell}(a) =\frac{1}{2} 
		\int_{-\frac{\ell}{2}}^{\frac{+\ell}{2}} \left(\mathsf{C_\varepsilon}(a) \varepsilon^2+2\mathsf{C_{\varepsilon\chi}}(a) \varepsilon\chi +\mathsf{C_{\chi}} (a)\chi^2 \right) \,\text{d} x_1,
		\label{eq:potential_energy_beam_dimensional_lc_forces_stiffness}
\end{equation}
where we have set
\begin{equation}
\varepsilon=w_{,1}, \quad \chi=v_{,11}
\end{equation}
for the axial strain and the curvature. The stiffness coefficients turn out to be:

\begin{equation}\label{stiffnesses}
	\begin{aligned}
		&\mathsf{C_\varepsilon}(a)=\frac{b h Y}{1-\nu^2}\,\frac{\left( 1+\frac{h}{\ell}\Phi^M(a) \right)}{\left( 1+\frac{h}{\ell}H_w^N(a) \right)\left(   1+\frac{h}{\ell}\Phi^M(a) \right)-\frac{1}{12}\left( \frac{h}{\ell}\Phi^N(a) \right)^2},   \\
		&\mathsf{C_{\varepsilon\chi}}(a)=\frac{12 b h^2 Y}{1-\nu^2}\,\frac{\frac{h}{\ell}\Phi^N(a)}{\left( 1+\frac{h}{\ell}H_w^N(a) \right)\left(   1+\frac{h}{\ell}\Phi^M(a) \right)-\frac{1}{12}\left( \frac{h}{\ell}\Phi^N(a) \right)^2},  \\
		&\mathsf{C_{\chi}}(a)=\frac{ b h^3 Y}{12(1-\nu^2)}\,\frac{\left( 1+\frac{h}{\ell}H^N_w(a) \right)}{\left( 1+\frac{h}{\ell}H_w^N(a) \right)\left(   1+\frac{h}{\ell}\Phi^M(a) \right)-\frac{1}{12}\left( \frac{h}{\ell}\Phi^N(a) \right)^2},
	\end{aligned}
\end{equation}
and the constitutive law for the beams therefore reads as in \eqref{constlaw}, plotted in Fig. \ref{fig:rigidezze}  for $\ell=h$.

\section{Results and discussion}
\begin{figure}
	\centering
	\includegraphics[scale=.6]{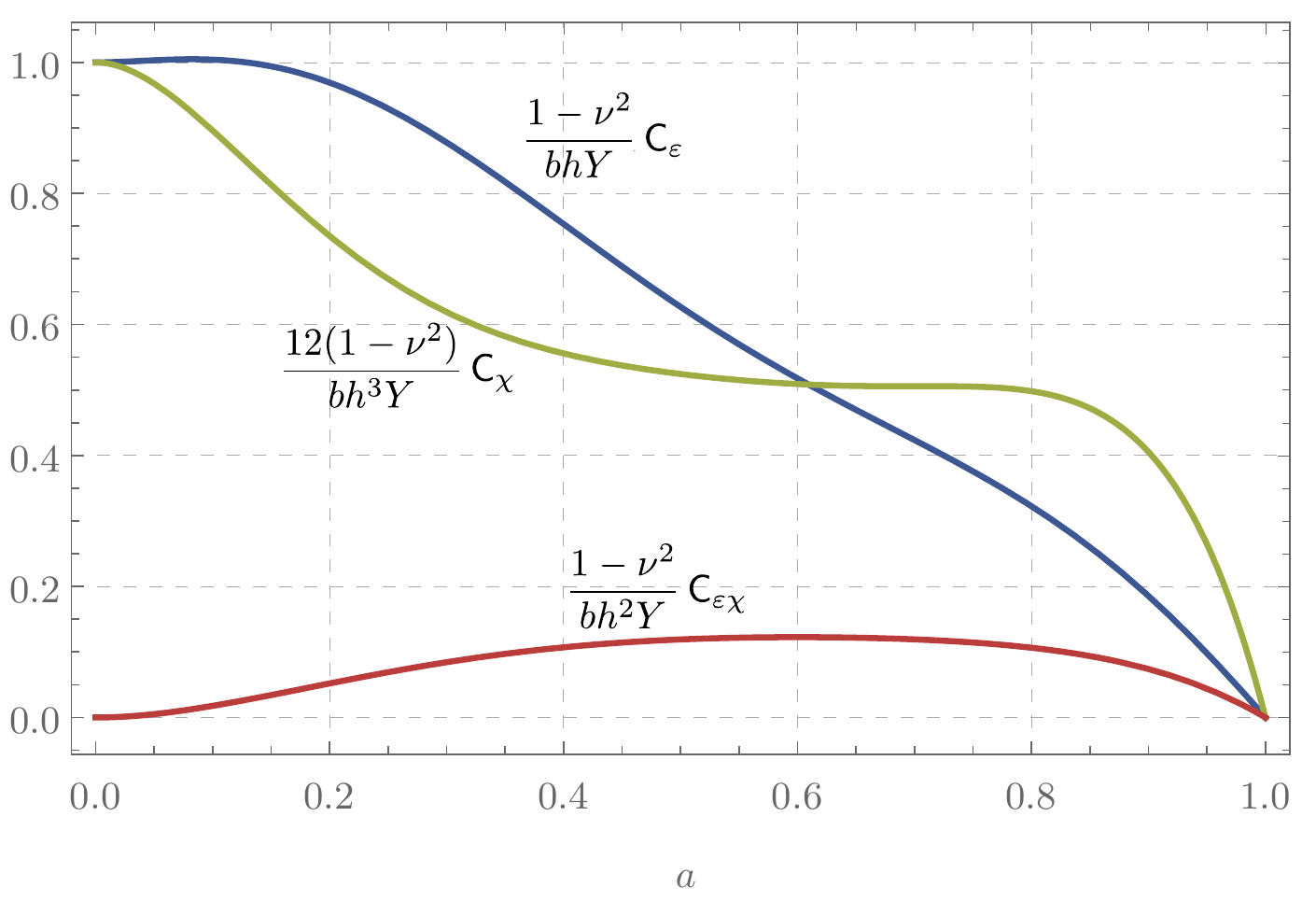}
	\caption{Stiffness coefficients \eqref{stiffnesses} as functions of $a$, the normalized depth of a crack stemming from the upper boundary of the beam (see Fig. \ref{fig:geometria}).}
	\label{fig:rigidezze}
\end{figure}

In Fig. \ref{fig:rigidezze} we reported the stiffness coefficients, determined through the procedure described in Sec. \ref{appendix-a-calculation-of-the-coefficients}.
Some comments are necessary:
\begin{enumerate}
	\item The presence of a crack, stemming from the upper boundary and having depth $a$, affects the stiffnesses in a different manner. As one would expect, the axial stiffness $\mathsf{C_\varepsilon}(a)$ decreases when the crack increases, essentially in a monotonic way and without relevant acceleration.  This reflects the fact that in a pure traction state, what really counts in the resistance mechanism is the effective height of the beam, whereas it does not matter if the crack is developed on the top/bottom or in the middle.

	The bending stiffness $\mathsf{C_\chi}(a)$ has instead a rapid decrease when the crack is initiated and, then, exhibits a plateau as the crack reaches the beam center line.
This can be explained because in a pure bending state the resistance mechanism is related to the moment of inertia: when a portion of materials goes missing at the top and the bottom of the beam, it dramatically decreases; instead, when the material is removed in the central part, the moment of inertia is not significantly affected.
	
	The coupling term $\mathsf{C_{\varepsilon\chi}}(a)$ vanishes for $a=0$, since the beam is supposed to be symmetric in the virgin state. When $0<a<1$ the symmetry is broken and this coupling stiffness increases up to a maximum value; for $a\to1$ the symmetry is recovered and then $\mathsf{C_{\varepsilon\chi}}$. The actual sign of  $\mathsf{C_{\varepsilon\chi}}$  depends on the positive signs assumed for $\varepsilon$ and $\chi$.
	\item As shown in Figs. \ref{fig:conv_Cb} and \ref{fig:rigidezze} and as proved in appendix \ref{slope}, both the compliances and stiffnesses scale as $O(a^2)$ for small cracks.  This feature has relevant consequences. The crack nucleation is possible only if the energy release rate evaluated at $a=0$, is greater (or equal) than the material toughness: $-\partial\Uc^\ell(a)/ \partial a\geq G_c>0$. 

%
%
	%
	Since in our case, $ \mathsf{C_\varepsilon}(a)$, $\mathsf{C_\chi}(a)$, and $\mathsf{C_{\varepsilon\chi}}(a)$ scale as $O(a^2)$, their derivatives  vanish at $a=0$; therefore, the energy \eqref{eq:potential_energy_beam_dimensional_lc_forces_stiffness}, endowed with a Griffith-like dissipation, has a local minimum at $a=0$ for all $\varepsilon$ and $\chi$.  Hence, this  model is not able to capture the crack nucleation \cite{Tanne_2018}.
	Including the presence of  cohesive forces between the crack lips could overcome this deficiency, but this is out of the scope of the present paper.

\item  
\begin{figure}[h!]
	\centering
	\includegraphics[scale=.4]{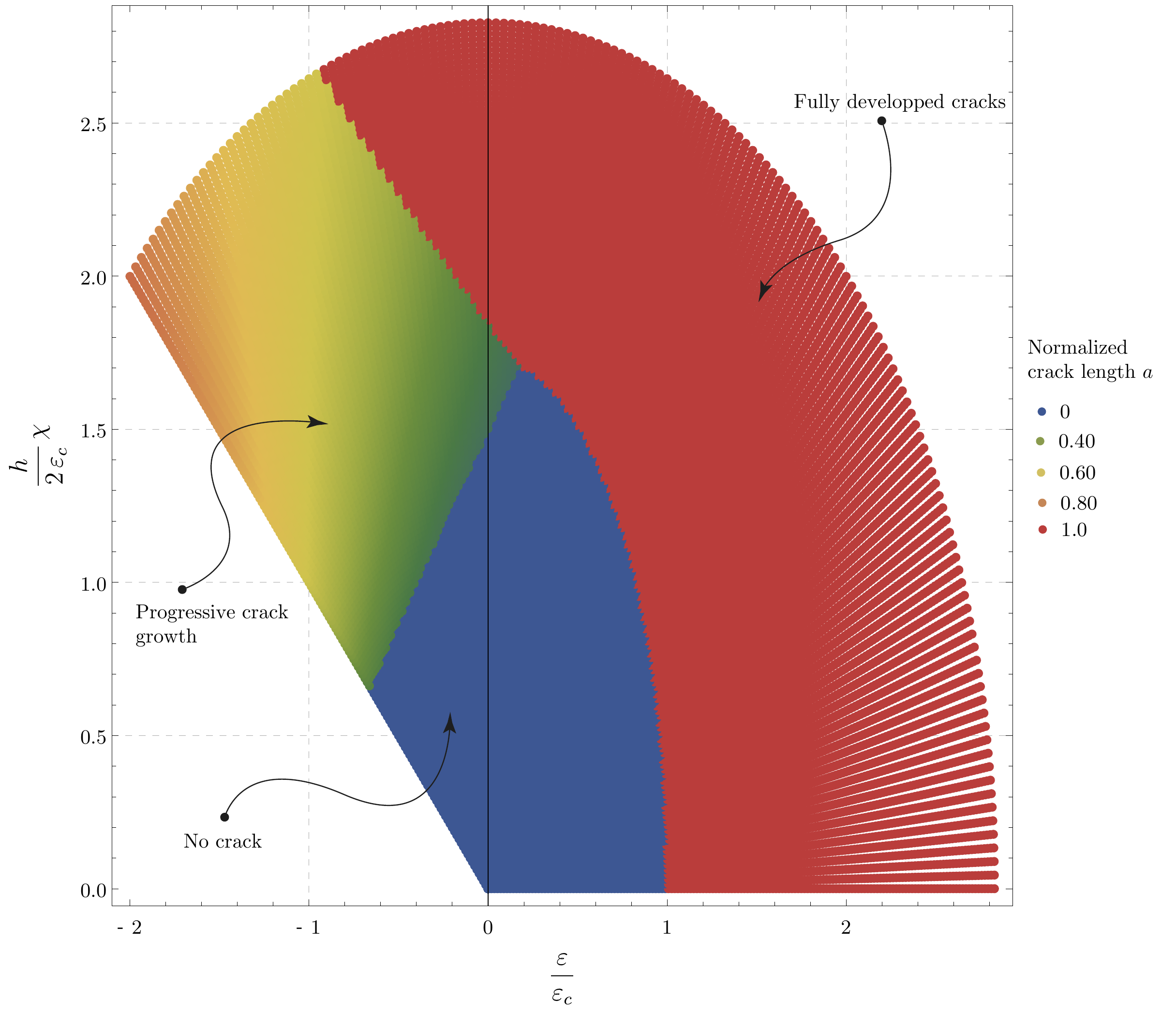}
	\caption{Global minima of \eqref{enfrac1d}: for all $\varepsilon$ and $\chi$, a color is associated to the global minimum of the functional.}
	\label{fig:ventaglio}
\end{figure}
As noted, $a=0$ is  always a minimum, be it  global o local; in the latter case other minima are possible. In Fig. \ref{fig:ventaglio}, for all $\varepsilon$ and $\chi$, we associate a colour to the global minimum of the functional
\begin{equation}
 \int_{-\frac{\ell}{2}}^{+\frac{\ell}{2}} 
	\Big( \mathsf{C_\varepsilon}(a) \frac{\varepsilon^2}{2} + \mathsf{C_\chi}(a) 
	\frac{\chi^2}{2} + \mathsf{C_{\varepsilon\chi}}(a)\varepsilon\chi + G_c h b a
	\Big) \text{d} x_1.
\end{equation}\label{enfrac1d}
The point $a=0$ is a global minimum just in a bounded neighborhood of the origin (blue points) . Out of this region, the global minima corresponds to  fully developed cracks ($a=1$, red points), or  partially developed cracks ($0<a<1$, gradient color scale).  If we consider a monotonic deformation path $(\varepsilon=\lambda\cos\varphi, \chi=\lambda\sin\varphi)$, the  coupling between axial and bending strains allows, not only for sudden fractures across the whole thickness, but also for more complex  evolution, where the crack growth can be progressive. These different scenarios are consistent with numerical evidence of Fig. \ref{fig:epschidiagram}, depend just on the ratio $\varepsilon/\chi=\tan\varphi$, and are made possible by the deduced constitutive equations \eqref{constlaw} and \eqref{stiffnesses}.

\end{enumerate}

We believe that it would be desirable to develop a phase-field gradient model including the present findings but also able to describe the crack nucleation, thanks to additional constitutive parameters, such as a characteristic length.

\appendix

\section{Non-dimensionalized bulk and boundary equations}\label{nondimeq}

Applying the non-dimensionalization introduced above to equations
(\ref{eq:elasticity_dimensional}), we obtain:
\begin{enumerate}
	\item  balance equations:
	\begin{equation}
		\eta S_{\alpha 1,1}^{(\eta)} + S_{\alpha 2,2}^{(\eta)} = 0,
		\label{eq:stresses_non_dimensional}
	\end{equation}
	\item constitutive equation (Voigt notation): 
	\begin{equation}  \label{eq:stresses_strains_non_dimensional}
		\begin{pmatrix}
			S_{11}^{(\eta)}\\
			S_{22}^{(\eta)}\\
			S_{12}^{(\eta)}
		\end{pmatrix} = 
		\frac{1}{(2\nu-1)(1+\nu)} \begin{pmatrix}
			\nu -1 & -\nu & 0 \\
			-\nu & \nu-1 & 0 \\
			0 & 0 & 2\nu-1 
		\end{pmatrix} \begin{pmatrix}
			E_{11}^{(\eta)}\\
			E_{22}^{(\eta)}\\
			E_{12}^{(\eta)}
		\end{pmatrix},
	\end{equation}
	%
	%
	%
	%
	%
	\item compatibility equation: 
	\begin{equation} \label{eq:compatibility_non_dimensional}
		\frac{\partial^2 E_{11}^{(\eta)}}{\partial y^2}
		+ \eta^2\frac{\partial^2 E_{22}^{(\eta)}}{\partial \bar{x}_1^2}
		= 2 \eta \frac{\partial^2 E_{12}^{(\eta)}}{\partial \bar{x}_1 \partial y},
	\end{equation}
	where 
	\begin{equation}
		E_{11}^{(\eta)} = \frac{1}{\eta} \frac{\partial u_1^{(\eta)}}{\bar{x}_1}, \qquad
		E_{22}^{(\eta)} = \frac{1}{\eta^2} \frac{\partial u_2^{(\eta)}}{\partial y}, \qquad
		2E_{12}^{(\eta)} = \frac{1}{\eta^2} \frac{\partial u_1^{(\eta)}}{\partial y} + \frac{1}{\eta} \frac{\partial u_2^{(\eta)}}{\partial \bar{x}_1}.
		\label{eq:strains_displacement_non_dimensional}
	\end{equation}
\end{enumerate} 

To these we must add the non-dimensional boundary conditions:

\begin{equation}
	\Sb^{(\eta)} \eb_2 = \mathbf{0}, \qquad \text{at } y = \pm 1/2,
	\label{eq:bc1_non_dimensional}
\end{equation}

\begin{equation}
	\Sb^{(\eta)} \eb_1 = (-12 M_o^{(\eta)} y + N_o^{(\eta)})\eb_1,
	\qquad \text{at } \bar{x}_1 = \pm 1/2,
	\label{eq:bc2_non_dimensional}
\end{equation}

\begin{equation}
	\int_{-1/2}^{1/2} S^{(\eta)}_{11}\dd y = N_o^{(\eta)},	\label{eq:bc3_non_dimensional}
\end{equation}

\begin{equation}
	-\int_{-1/2}^{1/2}y S^{(\eta)}_{11}\dd y = M_o^{(\eta)}, \qquad \text{at } \bar{x}_1 = \pm 1/2,
	\label{eq:bc4_non_dimensional}
\end{equation}
where the non-dimensional bending moment and traction at the boundary,
\(M_o^{(\eta)}\) and \(N_o^{(\eta)}\) are introduced. Their
non-dimensionalization follows the convention used in
(\ref{eq:non_dimensional_moments_and_tractions}).

\section{The outer problem}\label{outerappendix}

\paragraph{Order $-2$}

From the first equation of
(\ref{eq:stresses_non_dimensional}), it follows that
\(S^{(-2)}_{\alpha 2,2} = 0\). From the boundary condition
(\ref{eq:bc1_non_dimensional}), it follows that
\(S^{(-2)}_{\alpha 2} = 0\), so in particular also \(E^{(-2)}_{12} = 0\).
Since the expansion of the displacements starts at order \(0\), from
(\ref{eq:strains_displacement_non_dimensional}) it follows that also
\(E^{(-2)}_{11} = 0\). Then, from the constitutive equation, it follows that also
\(S^{(-2)}_{11} = 0\), and thus \(E^{(-2)}_{22} = 0\). Summing up, we have
\(\Sb^{(-2)} = \bm{0} = \Eb^{(-2)}\). Then, equation
(\ref{eq:strains_displacement_non_dimensional}) results in
\begin{equation}
	u^{(0)}_{1,2} = 0 = u^{(0)}_{2,2} \implies \ub^{(0)} = \ub^{(0)}(\bar{x}_1).
\end{equation}

\paragraph{Orded \(-1\)} From the first equation of
(\ref{eq:stresses_non_dimensional}), and the results at order \(-2\), it
still holds that \(S^{(-1)}_{\alpha 2,2} = 0\). From the boundary
condition (\ref{eq:bc1_non_dimensional}), it follows that
\(S^{(-1)}_{\alpha 2} = 0\), so in particular also \(E^{(-1)}_{12} = 0\).
From (\ref{eq:compatibility_non_dimensional}) and the results at order
\(-2\), it follows that
\(E^{(-1)}_{11,22} = 0 \implies E^{(-1)}_{11} = a^{(-1)}(\bar{x}_1) + b^{(-1)}(\bar{x}_1)y\).
Given that the first equation in
(\ref{eq:strains_displacement_non_dimensional}) relates \(E^{(-1)}_{11}\)
to \(\ub^{(0)}_1(\bar{x}_1)\), we see that necessarily
\(b^{(-1)}(\bar{x}_1) = 0\). Then, \(E^{(-1)}_{11} = a^{(-1)}(\bar{x}_1)\),
and
\(S^{(-1)}_{11} = E^{(-1)}_{11}/(1-\nu^2) = a^{(-1)}(\bar{x}_1)/(1-\nu^2)\).
From (\ref{eq:bc3_non_dimensional_simplified}), and the choice of the
scaling of external loads, it follows that \(a^{(-1)}(\bar{x}_1) = 0\).
The final result is

\begin{equation}
	\Eb^{(-1)} = \bm{0} = \Sb^{(-1)}.
	\label{eq:strains_and_stresses_outer_-1}
\end{equation}

Going back to \(\ub^{(0)}(\bar{x}_1)\), again the first equation of
(\ref{eq:strains_displacement_non_dimensional}) gives

\begin{equation}
	u^{(0)}_{1,1}(\bar{x}_1) = a^{(-1)}(\bar{x}_1) = 0 \implies u^{(0)}_1 = 0
	\implies \ub^{(0)} = v^{(0)}(\bar{x}_1)\eb_2,
	\label{eq:u0_outer}
\end{equation}

where no constant of integration appears since {we assume that
	the center of mass of the base \(\bar{x}_1 = -\frac{1}{2}\) is blocked.

From (\ref{eq:strains_displacement_non_dimensional}) at the next order
the condition on \(\ub^{(1)}\) is \(u^{(1)}_{2,2} = E^{(-1)}_{22} = 0\) and
\(v^{(0)}_{,1} + u^{(1)}_{1,2} = 0\). These two imply that

\begin{equation}
	\ub^{(1)} =\left( w^{(1)}(\bar{x}_1) - y v^{(0)}_{,1}(\bar{x}_1)\right) \eb_1 +
	v^{(1)}(\bar{x}_1) \eb_2.
\end{equation}

\paragraph{Order \(0\)}

From the first equation of
(\ref{eq:stresses_non_dimensional}), the boundary condition
(\ref{eq:bc1_non_dimensional}) and the results at order \(-1\), it still
holds that \(S^{(0)}_{\alpha 2} = 0\), so in particular also
\(E^{(0)}_{12} = 0\), and \(E^{(0)}_{22}\), \(E^{(0)}_{11}\) are multiple of each
other. From (\ref{eq:compatibility_non_dimensional}) and the results at
order \(-1\), it follows that
\(E^{(0)}_{11,22} = 0 \implies E^{(0)}_{11} = a^{(0)}(\bar{x}_1) + b^{(0)}(\bar{x}_1)y\).
Going back to the expression of \(\ub^{(1)}\) obtained previously, and
using the first equation of
(\ref{eq:strains_displacement_non_dimensional}), it follows that
\(b^{(1)}(\bar{x}_1) = -v^{(0)}_{,11}(\bar{x}_1)\), and that
\(w^{(1)}(\bar{x}_1) = A^{(0)}(\bar{x}_1)\), where \(A^{(0)}\) is the integral of
\(a^{(0)}(\bar{x}_1)\). From boundary conditions
(\ref{eq:bc3_non_dimensional_simplified}) at this order

\begin{equation}
	\int_{-1/2}^{1/2} S^{(0)}_{11} \text{d}y = \bar{N} \implies \frac{a^{(0)}(\bar{x}_1)}{1-\nu^2} =
	\bar{N} \implies a^{(0)}(\bar{x}_1) = w^{(1)}_{,1}(\bar{x}_1) = (1-\nu^2) \bar{N},
\end{equation}

and

\begin{equation}
	S^{(0)}_{11} =  \frac{w^{(1)}_{,1}(\bar{x}_1) - y v^{(0)}_{,11}(\bar{x}_1)}{1-\nu^2}.
	\label{eq:s0_outer_general}
\end{equation}

Then,

\begin{equation}
	E^{(0)}_{11} = u^{(1)}_{1,1} = w^{(1)}_{,1}(\bar{x}_1) -v^{(0)}_{,11}(\bar{x}_1)y,
\end{equation}

and from these, using (\ref{eq:stresses_strains_non_dimensional},
\ref{eq:strains_stresses_non_dimensional}), the other quantities can be
deduced

\begin{equation}
	E^{(0)}_{22} = -\nu(1+\nu) S^{(0)}_{11} = \frac{-\nu}{1-\nu}
	\left(w^{(1)}_{,1}(\bar{x}_1) - v^{(0)}_{,11}(\bar{x}_1)y\right).
	\label{eq:e0_22_outer_general}
\end{equation}

The expression of \(\ub^{(1)}\) is then \begin{equation}
	\begin{aligned}
		\ub^{(1)} &= \left(w^{(1)}(\bar{x}_1) - y v^{(0)}_{,1}(\bar{x}_1) \right)\eb_1
		+ v^{(1)}(\bar{x}_1)\eb_2, \\
	\end{aligned}
	\label{eq:u1_outer}
\end{equation}

Using as before the relations
(\ref{eq:strains_displacement_non_dimensional}) between strains and
displacements the next order of the displacements is calculated,
resulting in

\begin{equation}
	\begin{aligned}
		u^{(2)}_1 &= w^{(2)}(\bar{x}_1) -y v^{(1)}_{,1}(\bar{x}_1), \\
		u^{(2)}_2 &= v^{(2)}(\bar{x}_1) -\frac{\nu}{1-\nu}
		\left(w^{(1)}_{,1}(\bar{x}_1)y - v^{(0)}_{,11}(\bar{x}_1) \frac{y^2}{2}\right).
	\end{aligned}
	\label{eq:u2_outer_general}
\end{equation}

\paragraph{Order \(1\)}

From the first equation of
(\ref{eq:stresses_non_dimensional}), at this order two different
conditions are satisfied:

\begin{equation}
	S^{(0)}_{\alpha 1,1} = -S^{(1)}_{\alpha 2,2};
\end{equation}
from these, the results at previous order and boundary conditions we
still have \(S^{(1)}_{22} = 0\), while now
\begin{equation}
	S^{(1)}_{12,2} = -S^{(0)}_{11,1} = \frac{-w^{(1)}_{,11}(\bar{x}_1) + y v^{(0)}_{,111}(\bar{x}_1)}{1-\nu^2}.
\end{equation}

From (\ref{eq:compatibility_non_dimensional}) and the results at order
\(0\), it follows that
\(E^{(1)}_{11,22} = 0 \implies E^{(1)}_{11} = a^{(1)}(\bar{x}_1) + b^{(1)}(\bar{x}_1)y\).
Going back to the general expression of \(\ub^{(2)}\)
(\ref{eq:u2_outer_general}), and using the first equation of
(\ref{eq:strains_displacement_non_dimensional}), it follows that
\begin{equation}
	E^{(1)}_{11} = u^{(2)}_{1,1} = w^{(2)}_{,1}(\bar{x}_1) -y v^{(1)}_{,11}(\bar{x}_1),
\end{equation}
therefore \(b^{(1)}(\bar{x}_1) = -v^{(1)}_{,11}(\bar{x}_1)\), and
\(w^{(2)}(\bar{x}_1) = A^{(1)}(\bar{x}_1)\), where \(A^{(1)}\) is the integral of
\(a^{(1)}(\bar{x}_1)\). The expressions for the stresses at this order are
\begin{equation}
	\begin{aligned}
		S^{(1)}_{11} &= \frac{E^{(1)}_{11}}{1-\nu^2} = 
		\frac{w^{(2)}_{,1}(\bar{x}_1) - yv^{(1)}_{,11}(\bar{x}_1)}{1-\nu^2},\\
		S^{(1)}_{12} &= \frac{4y^2-1}{8(1-\nu^2)} v^{(0)}_{,111}(\bar{x}_1),
	\end{aligned}
	\label{eq:s1_outer_general}
\end{equation}

while the strains read

\begin{equation}
	\begin{aligned}
		E^{(1)}_{11} &= w^{(2)}_{,1}(\bar{x}_1) -y v^{(1)}_{,11}(\bar{x}_1),\\
		E^{(1)}_{12} &= (1+\nu)S^{(1)}_{12} = \frac{4y^2-1}{8(1-\nu)} v^{(0)}_{,111}(\bar{x}_1),\\
		E^{(1)}_{22} &= u^{(3)}_{2,2} = -\frac{\nu}{1-\nu}
		\left(w^{(2)}_{,1}(\bar{x}_1) - y v^{(1)}_{,11}(\bar{x}_1)\right).
	\end{aligned}
	\label{eq:strain_order_1_general}
\end{equation}

\section{The inner problem}\label{innerappendix}
Given the rescaling of the \(\bar{x}_1\) coordinate, the equations of
elasticity to be solved for the inner problem no longer show coupling between the
different orders:

-balance equations: 
\begin{equation}
	\mathsf{S}_{\alpha 2,2}^{(\eta)} + \mathsf{S}_{\alpha 1,1}^{(\eta)} = 0,
	\label{eq:stresses_non_dimensional_inner}
\end{equation}


	-compatibility equation:
	
	\begin{equation}
		\frac{\partial^2 \mathsf{E}_{11}^{(\eta)}}{\partial x}
		+ \frac{\partial^2 \mathsf{E}_{22}^{(\eta)}}{\partial y^2}
		= 2 \frac{\partial^2 \mathsf{E}_{12}^{(\eta)}}{\partial x \partial y},
		\label{eq:compatibility_non_dimensional_inner}
	\end{equation}
	with 
	\begin{equation}
		\mathsf{E}_{11}^{(\eta)} = \frac{1}{\eta^2} \frac{\partial v_1^{(\eta)}}{\partial y}, \qquad
		\mathsf{E}_{22}^{(\eta)} = \frac{1}{\eta^2} \frac{\partial v_2^{(\eta)}}{\partial x}, \qquad
		2\mathsf{E}_{12}^{(\eta)} = \frac{1}{\eta^2} \frac{\partial v_1^{(\eta)}}{\partial x} + \frac{1}{\eta^2} \frac{\partial v_2^{(\eta)}}{\partial y}.
		\label{eq:strains_displacement_non_dimensional_inner}
	\end{equation}

\paragraph{Order \(-2\)} 
We have \(\div\, \Ssb^{(-2)} = \bm{0}\), with
\(\lim_{x\to\pm\infty}\Ssb^{(-2)}(x, y) = \bm{0}\) from the matching
conditions, and a stress free condition on the other boundaries.
Multiplying (\ref{eq:stresses_non_dimensional_inner}) by \(\usb^{(0)}\)
and integrating over the domain gives

\begin{equation}
	\int_{\Omega} \div\Ssb^{(-2)}\cdot\usb^{(0)} \text{d}x\text{d}y = 0.
\end{equation}

Integrating by parts the latter equation, and taking boundary conditions
into account, results in

\begin{multline}
	\int_{\Omega} \div\Ssb^{(-2)}\cdot\usb^{(0)} \text{d}x\text{d}y =
	-\int_{\Omega} \Ssb^{(-2)}\cdot \nabla \usb^{(0)} \text{d}x\text{d}y + \int_{\partial\Omega} \Ssb^{(-2)} \nb \cdot\usb^{(0)} \text{d}l \\
	= -\int_{\Omega} \Ssb^{(-2)}\cdot \Esb^{(-2)} \text{d}x\text{d}y = 0.
\end{multline}

Using (\ref{eq:stresses_strains_non_dimensional}), it follows that
\(\Esb^{(-2)}(\usb^{(0)}) = \mathbf{0}\), thus \(\Ssb^{(-2)} =  \mathbf{0}\). This
last result implies that the field \(\usb^{(0)}\) is a rigid displacement:

\begin{equation}
	\usb^{(0)}(x, y) = \tb^{(0)} + \omega^{(0)}\eb_3 \times \yb,
	\label{eq:rigid_displacement_0}
\end{equation}
where we introduced the vector \(\yb =x\eb_1+y\eb_2\), and \(\tb^{(0)}\),
\(\omega^{(0)}\), both constant, are a translation and a rotation, respectively. Since from the outer
expansion we know that \(\ub^{(0)} = v^{(0)}(\bar{x}_1)\eb_2\), it follows
that the \(\eb_1\) component of (\ref{eq:rigid_displacement_0}) must
vanish

\begin{equation}
	t^{(0)}_1 - \omega^{(0)} y  = 0 \qquad \forall y,
\end{equation}
which in turn implies \(\omega^{(0)} = 0 = t^{(0)}_1\), and
\(v^{(0)}(0^\pm) = t^{(0)}_2\), from which the continuity of \(v^{(0)}\) at \(0\)
follows:

\begin{equation}
	[\![ v^{(0)} ]\!] = 0 \qquad \text{at } \bar{x}_1 = 0,
	\label{eq:continuity_u_0}
\end{equation}
where the jump at a specific value of \(\bar{x}_1\) is defined as
\([\![ f ]\!] = f(\bar{x}_1^+)-f(\bar{x}_1^-)\).

\paragraph{Order \(-1\)} 

We have \(\div \Ssb^{(-1)} = \mathbf{0}\), and repeating
the same procedure used for order \(-2\) yields
\(\Esb^{(-1)}(\usb^{(1)}) = \mathbf{0}= \Ssb^{(-1)}\). This in turn
implies that the displacement field at order \(-1\) is also rigid

\begin{equation}
	\usb^{(1)}(x, y) = \tb^{(1)} + \omega^{(1)}\eb_3 \times \yb.
	\label{eq:rigid_displacement_1}
\end{equation}

Using conditions (\ref{eq:u0_outer}, \ref{eq:u1_outer}) together with
the second of the matching conditions (\ref{eq:u_matching_conditions}),
it follows that:

\begin{multline}
	\tb^{(1)} - \omega^{(1)} y \eb_1 + \lim_{x\to\pm\infty}
	\omega^{(1)} x \eb_2 = \\
	\left[w^{(1)}(0^\pm) - yv^{(0)}_{,1}(0^\pm)\right]\eb_1 +
	v^{(1)}(0^\pm)\eb_2 + \lim_{x\to\pm\infty}
	x v^{(0)}_{,1}(0^\pm)\eb_2,
\end{multline}
from which we have that necessarily \(\omega^{(1)} = v^{(0)}_{,1}(0^\pm)\), and
\(t^{(1)}_1 = w^{(1)}(\bar{x}_1)\), \(t^{(1)}_2 = v^{(1)}(0^\pm)\). This implies the
continuity for the three quantities
\begin{equation}
	[\![ v^{(1)} ]\!] = [\![ v^{(0)}_{,1} ]\!] = [\![ w^{(1)} ]\!] = 0
	\qquad \text{at } \bar{x}_1 = 0.
	\label{eq:continuity_u_1}
\end{equation}
and the expression of \(\usb^{(1)}\) is determined:

\begin{equation}
	\usb^{(1)}(x, y) =  \left(w^{(1)}(0) - y v^{(0)}_{,1} (0) \right) \eb_1 + 
	\left(x v^{(0)}_{,1} (0) + v^{(1)}(0) \right) \eb_2
	\label{eq:v1_expr}
\end{equation}

\paragraph{Order $0$} 
The problem involving the next terms of the expansion is

\begin{equation}
	\begin{cases}
		&   \div\Ssb^{(0)} = \bm{0},\\
		&  \Esb^{(0)}  = -\nu \text{tr}\Ssb^{(0)} \Ib
		+ (1 + \nu) \Ssb^{(0)},\\
		&  \Ssb^{(0)} \nb = \bm{0} \hfill \text{on } \partial \Omega^+,\\
		&   \displaystyle\lim_{x\to\pm\infty} \Ssb^{(0)}(x, y) = \Sb^{(0)}(0^\pm, y). 
	\end{cases}
	\label{eq:inner_problem_order_0}
\end{equation}

The last equation of the problem comes from the matching conditions and
we recall that \(S^{(0)}_{12} = S^{(0)}_{22} = 0\), while
\(S^{(0)}_{11}(0^\pm) = \frac{1}{1-\nu^2}\, \big(w^{(1)}_{,1}(0^\pm)-yv^{(0)}_{,11}(0^\pm)\big)\).
Furthermore, the third matching condition in
(\ref{eq:u_matching_conditions}) applies. We want to allow for rigid
motions in the solution of the problem (\ref{eq:inner_problem_order_0}),
so we introduce the condition that

\begin{equation}
	\int_{\partial\Omega} \yb \times \Ssb^{(0)} \nb 
	\cdot\eb_3 \text{d}l = 0 \implies \int_{-1/2}^{1/2} y S^{(0)}_{11}(0^+, y)\dd y 
	= \int_{-1/2}^{1/2} y S^{(0)}_{11}(0^-, y)\dd y,
\end{equation}
where we also used that \(S^{(0)}_{12} = 0\). It is clear that this
condition implies that the bending moment be continuous at
\(\bar{x}_1 = 0\), that is
\begin{equation}
	[\![ v^{(0)}_{,11} ]\!] = 
	[\![ M^{(0)} ]\!] = 0 \qquad \text{at } \bar{x}_1 = 0.
	\label{eq:continuity_M_0_11}
\end{equation}

It follows from (\ref{eq:bending_moments_order_0_1}) that \(v^{(0)}_{,11}\)
is also continuous at \(\bar{x}_1 = 0\). Imposing the further condition

\begin{multline}
	\int_{\partial\Omega} \Ssb^{(0)} \nb 
	\cdot\eb_1 \text{d}l = 0 \implies \int_{-1/2}^{1/2} S^{(0)}_{11}(0^+, y)\dd y 
	= \int_{-1/2}^{1/2} S^{(0)}_{11}(0^-, y)\dd y \\
	\implies \int_{-1/2}^{1/2} w^{(1)}_{,1}(0^+)\dd y = 
	\int_{-1/2}^{1/2} w^{(1)}_{,1}(0^-)\dd y,
\end{multline}
from which we have the transmission condition on \(w^{(1)}_{,1}\) and
\(N^{(0)}\) at \(\bar{x}_1 = 0\):
\begin{equation}
	[\![ w^{(1)}_{,1} ]\!] = 0 \implies [\![N^{(0)}]\!] = 0 
	\qquad \text{at } \bar{x}_1 = 0.
	\label{eq:continuity_W_1_1}
\end{equation}

	\vspace{1cm}
	
To get \eqref{eq:alternative_transmission_conditions_order_0} it is  necessary to substitute into the third equation of
	(\ref{eq:u_matching_conditions}), \eqref{eq:diff_Cd_Kd}  and the
	expressions of \(\ub\) up to second order (\ref{eq:u0_outer},
	\ref{eq:u1_outer}, \ref{eq:u2_outer_general}):
	
	\begin{multline}
		\lim_{x\to\pm\infty} (t^{(2)}_1 -\omega^{(2)} y + w^{(1)}_{,1}x -xy v^{(0)}_{,11} + 
		w^{(1)}_{,1}t^N_1 + v^{(0)}_{,11}t^M_1)\eb_1 \\ + \lim_{x\to\pm\infty} 
		\Bigg( \Bigg. 
		t^{(2)}_2 + \omega^{(2)}x -\frac{\nu}{1-\nu}w^{(1)}_{,1}y + 
		\left(\frac{\nu y^{(2)}}{2(1-\nu)} + \frac{x^{(2)}}{2}\right) v^{(0)}_{,11}\\
		+ w^{(1)}_{,1}t^N_2 + v^{(0)}_{,11}t^M_2  
		\Bigg. \Bigg) \eb_2 =\\
		\Bigg[ \Bigg. 
		(w^{(2)}(0^\pm) -yv^{(1)}_{,1}(0^\pm))\eb_1 + \left(v^{(2)}(0^\pm) 
		-\frac{\nu}{1-\nu}w^{(1)}_{,1}y 
		+ \frac{\nu v^{(0)}_{,11}y^{(2)}}{2(1-\nu)}\right)\eb_2 
		\Bigg. \Bigg] \\
		+ \lim_{x\to\pm\infty}\Bigg[ \Bigg. 
		\left(w^{(1)}_{,1}x -xy v^{(0)}_{,11}\right)
		\eb_1 + \left(x v^{(1)}_{,1}(0^\pm) + \frac{x^{(2)}}{2} v^{(0)}_{,11}\right)
		\eb_2\Bigg. \Bigg],
		\label{eq:matching_limit_general_inner_0}
	\end{multline}
	where again, continuous outer functions are implicitly evaluated at
	\(\bar{x}_1 = 0\), and the dependencies of the auxiliary functions on
	\(x, y\) have been left implicit. Simplifying:
	\begin{multline}
		\lim_{x\to\pm\infty}[(t^{(2)}_1 -\omega^{(2)} y +  w^{(1)}_{,1}t^N_1 + v^{(0)}_{,11}t^M_1)
		\eb_1 \\+ (t^{(2)}_2 + \omega^{(2)}x + w^{(1)}_{,1}t^N_2 + v^{(0)}_{,11}t^M_2)
		\eb_2] =\\
		[\left(w^{(2)}(0^\pm)-yv^{(1)}_{,1}(0^\pm)\right)\eb_1 + v^{(2)}(0^\pm) \eb_2]\\  
		+\lim_{x\to\pm\infty} x v^{(1)}_{,1}(0^\pm) \eb_2,
		\label{eq:matching_limit_general_inner_0_simplified}
	\end{multline}
	
	Gathering the terms with the same scaling, and recalling equations
	(\ref{eq:v_t_at_infinity_inner_0}, \ref{eq:v_b_at_infinity_inner_0}), it
	follows that
	
	\begin{equation}
		\begin{aligned}
			-\omega^{(2)}y - (\Phi^M)^{\pm}v^{(0)}_{,11}(0)y - (\Phi^N)^{\pm}w^{(1)}_{,1}(0)y
			&= -v^{(1)}_{,1}(0^\pm)y, \\
			t^{(2)}_1 + (H_w^N)^{\pm}w^{(1)}_{,1}(0) + (H^M_w)^{\pm}v^{(0)}_{,11}(0) 
			&= w^{(2)}(0^\pm), \\
			t^{(2)}_2 + (H_v^N)^{\pm}w^{(1)}_{,1}(0) + (H^M_v)^{\pm}v^{(0)}_{,11}(0) 
			&= v^{(2)}(0^\pm),
		\end{aligned}
		\label{eq:matching_conditions_order_0}
	\end{equation}
	resulting in the transmission conditions for \(v^{(1)}_{,1}\), \(v^{(2)}\) and
	\(w^{(2)}\) at \(\bar{x}_1 = 0\):
	
	\begin{equation}
		\begin{aligned}
			{}[\![v^{(1)}_{,1}]\!] &= \Phi^N w^{(1)}_{,1}(0) + \Phi^M v^{(0)}_{,11}(0), \\
			{}[\![w^{(2)}]\!] &= H_w^Nw^{(1)}_{,1}(0) + H^M_wv^{(0)}_{,11}(0), \\
			{}[\![v^{(2)}]\!] &= H_v^Nw^{(1)}_{,1}(0) + H^M_vv^{(0)}_{,11}(0).
		\end{aligned}
		\label{eq:transmssion_conditions_order_0}
	\end{equation}
	
	An equivalent relation to the first equation of
	(\ref{eq:matching_conditions_order_0}) can be obtained involving \(x\)
	in place of \(y\), yielding the same transmission condition for
	\(v^{(1)}_{,1}\). From (\ref{eq:bending_moments_order_0_1}),
	(\ref{eq:tractions_order_0_1}), we then obtain \eqref{eq:alternative_transmission_conditions_order_0}.
	
\section{Behaviour of the stiffness coeffficients for small cracks}\label{slope}

In order to prove that all the stiffnesses have a null slope at $a=0$, we first observe that	if 	the forces acting on an elastic body are confined to  distinct portions of its surface, 
each lying within a sphere of radius $a$, then the stresses and strains at a fixed
interior point of the body are of a smaller order of magnitude in $a$ as  $a\to 0$ \cite{Sternberg_1954}. We then conclude that $\Sigmab^N$ and $\Sigmab^M$ are at most of the order $a$:
\begin{equation}
	\Sigmab^N=a\Sigmab^N_1 + o(a), \quad \Sigmab^M=a\Sigmab^M_1+ o(a),
\end{equation}
where $\Sigmab^N_1$ and $\Sigmab^M_1$ are  independent of $a$. Then, in the light of \eqref{TM} and \eqref{TN}, we may write
\begin{equation}\label{TJ}
	\Tb^N=	\Tb^N_o+a \Tb^N_1+ o(a), \quad \Tb^M=	\Tb^M_o+a \Tb^N_1+ o(a),
\end{equation}
with $\Tb^J_o$ and $\Tb^J_1$  ($J=N,M$) independent of $a$.  Let us denote by $\overline\Sb$ the mean stress corresponding to a stress field $\Sb$, such that $\Sb$ is equilibrated, \textit{i.e.}, $\Sb$ satisfies the equilibrium equation $\div\Sb=\mathbf{0}$ in $\overline{\Omega}$;  $\overline{\Sb}$ depends only on the surface traction $\cb$ as follows\footnote{To prove this statement, let us consider a regular field $\wb$ in $\overline{\Omega}$; from the divergence theorem we get the Signorini identity (see Sect. 18 of \cite{Gurtin_1973}):
	\begin{equation}
		\int_{\partial\Omega}\wb\otimes\cb=	\int_{\partial\Omega}\wb\otimes\Sb\nb=\int_{\Omega}\wb\otimes\div\Sb+\int_{\Omega}(\nabla w)\Sb^T.
	\end{equation}
	On setting $\wb=\xb$ and making use of the equilibrium equation, we obtain \eqref{meanS}.
}:
\begin{equation}\label{meanS}
	\overline{\Sb}=\frac{1}{|\Omega|}\int_{\partial\Omega}\xb\otimes\cb.
\end{equation}
On denoting with  $\Gamma^\pm$ the  lips of the crack, from \eqref{meanS} we have that
\begin{equation}
	\overline{\Sigmab}^N=\int_{\Gamma^+}y\Sigma^N_{11}\eb_2\otimes\eb_1-\int_{\Gamma^-}y\Sigma^N_{11}\eb_2\otimes\eb_1=\mathbf{0};
\end{equation}
analogous conclusion is achieved for $\Sigmab^M$:
\begin{equation}
	\overline{\Sigmab}^M=\int_{\Gamma^+}y^2\Sigma^M_{11}\eb_2\otimes\eb_1-\int_{\Gamma^-}y^2\Sigma^M_{11}\eb_2\otimes\eb_1=\mathbf{0}.
\end{equation}
The result is a consequence of the fact that the applied traction on $\Gamma^\pm$ are  in astatic equilibrium\footnote{A system of external forces is said to be in \textit{astatic equilibrium} if it 	
	remains in equilibrium when turned through an arbitrary angle.} \cite{Gurtin_1973,Villaggio_1977}.
When determining the compliance coefficients  \eqref{compliance}, we therefore need to evaluate integrals as follows:
\begin{equation}
	\begin{aligned}
		&	\int_{\Omega}(\Tb_o^J+a\Tb_1^J+o(a))\cdot\Co(\Sigmab_o^K+a \Sigmab_1^K+o(a))=\\
		&	\Co\Tb_o^J\cdot\int_{\Omega}\Sigmab_o^K+a\left(\Co\Tb_o^J\cdot\int_{\Omega}\Sigmab_1^K+\Co\Tb_1^J\cdot\int_{\Omega}\Sigmab_o^K\right)+O(a^2) 
	\end{aligned}
\end{equation}
for $J,K=N,M$. Since the $\overline{\Sigmab}^K_o=\overline{\Sigmab}^K_1=0$, we conclude that the compliance coefficients are at most quadratic in $a$, which is sufficient to claim that their slope at $a=0$ is null.

\bibliographystyle{elsarticle-num}
\bibliography{biblio}
\addcontentsline{toc}{section}{References}

\end{document}